%% file: main.tex
\documentclass[sigconf]{acmart}

\acmConference[Preprint]{ACM Conference}{August 2024}{Houston,  Texas, USA}%

\usepackage[utf8]{inputenc} % allow utf-8 input
\usepackage[T1]{fontenc}    % use 8-bit T1 fonts
\usepackage{hyperref}       % hyperlinks
\usepackage{url}            % simple URL typesetting
\usepackage{booktabs}       % professional-quality tables
\usepackage{amsfonts}       % blackboard math symbols
\usepackage{nicefrac}       % compact symbols for 1/2, etc.
\usepackage{microtype}      % microtypography
\usepackage{lipsum}
\usepackage{fancyhdr}       % header
\usepackage{graphicx}       % graphics
\usepackage{subfig}
\usepackage{comment}
\usepackage{multicol}
\usepackage{multirow}
\usepackage{todonotes}
\usepackage{amsmath} 
\usepackage{cleveref}
\usepackage{fancybox}
\usepackage{placeins} 
\usepackage{algorithm}
\usepackage{algpseudocode}
\usepackage{xspace}
\usepackage[most]{tcolorbox} % Load tcolorbox package

\cornersize{.2}
\newcounter{observation}

\renewcommand\footnotetextcopyrightpermission[1]{} % Removes the footnote with conference information
\fancypagestyle{plain}{
    \fancyhf{} % Clear all header and footer fields
\fancyhead[R]{Your Right Header} % Change 'Your Right Header' to your desired right header text
}
\newcommand{\tool}{\textsc{Lya}\xspace}

\title{Unlearning Trojans in Large Language Models: A Comparison Between Natural Language and Source Code}

\author{Mahdi Kazemi}
\affiliation{%
  \institution{University of Houston}
  \city{Houston, Texas}
  \country{USA}
}
\email{mahdikazemi76@gmail.com }

\author{Aftab Hussain}
\affiliation{%
  \institution{University of Houston}
  \city{Houston, Texas}
  \country{USA}
}
\email{ahussain27@uh.edu}

\author{Md Rafiqul Islam Rabin}
\affiliation{%
  \institution{University of Houston}
  \city{Houston, Texas}
  \country{USA}
}
\email{mrabin@uh.edu}

\author{Mohammad Amin Alipour}
\affiliation{%
  \institution{University of Houston}
  \city{Houston, Texas}
  \country{USA}
}
\email{maalipou@central.uh.edu }

\author{Sen Lin}
\affiliation{%
  \institution{University of Houston}
  \city{Houston, Texas}
  \country{USA}
}
\email{slin50@central.uh.edu }

\begin{document}
\input{abstract}


\maketitle

\input{intro}

\input{related}

\input{methodology}

\input{setup}

\input{results}

\input{disc-concl}

\bibliographystyle{ACM-Reference-Format}
\bibliography{unlearning}

\end{document}

%% file: abstract.tex
\begin{abstract}

This work investigates the application of Machine Unlearning (MU) for mitigating the impact of trojans embedded in conventional large language models of natural language (Text-LLMs) and large language models of code (Code-LLMs) We propose a novel unlearning approach, \tool, that leverages both gradient ascent and elastic weight consolidation, a Fisher Information Matrix (FIM) based regularization technique, to unlearn trojans from poisoned models. We compare the effectiveness of \tool against conventional techniques like fine-tuning, retraining, and vanilla gradient ascent. The subject models we investigate are BERT and CodeBERT, for sentiment analysis and code defect detection tasks, respectively. Our findings demonstrate that the combination of gradient ascent and FIM-based regularization, as done in \tool, outperforms existing methods in removing the trojan's influence from the poisoned model, while preserving its original functionality. To the best of our knowledge, this is the first work that compares and contrasts MU of trojans in LLMs, in the NL and Coding domain.

%\textcolor{red}{`Comment for Mahdi/Aftab: only LLM for natural language and Code-LLM for source code? I have seen in several places you used only LLMs to refer to both natural and code. Better to use something NL-LLM or Text-LLM to specify natural language LLM? <Please discard if I misunderstood.>'}. -- seems we clarify what we mean by LLMs in each context that we use. e.g. when we use generally we say LLMs in the NL and Coding domain. -- in the abstract, I added Text-LLMs

%DONE \textcolor{red}{`Comment for Mahdi/Aftab: the abbreviation FIM has not been defined yet'}

\end{abstract}

%% file: intro.tex
\section{Introduction}
\label{sec:intro}

%LLMs
Large Language Models (LLMs) are increasingly being utilized in software development. These models can generate code snippets, provide auto-completion suggestions, or even assist in writing documentation. By leveraging the vast amount of data on which they have been trained, LLMs can aid developers in various tasks, accelerating the software development process and improving overall productivity.

LLMs internalize different knowledge from the training data to produce output for different application and different domains. Depending on the training dataset and the training process, the output can contain undesired contents. For example, in a natural language chat application, the output may contain harmful language or misinformation, or in the domain of software engineering, it may contain vulnerable code.
Unfortunately, the process of data collection and training is very expensive. Therefore, it is impractical to retrain a new LLM once some bad behavior is detected in the model. Therefore, one of emerging fields are editing the knowledge contents of LLMs to modify the behavior of the models after training; in this case, unlearning the undesired behavior.

Model unlearning, also referred to as machine unlearning (MU), is the process making a machine learning model forget specific information it has learned during training~\cite{bourtoule2021machine}. Model unlearning is therefore essentially the mirror-opposite of machine learning, where the goal is to expunge certain data or behaviors from a model without retraining it from scratch, and thereby avoids the high computation costs of retraining. This approach is helpful when we want to address privacy concerns by allowing models to forget sensitive, undesirable, or deprecated information. However, unlearning is challenging due to the complex and stochastic nature of training methods~\cite{bourtoule2021machine}, and the risk of a phenomenon similar to catastrophic forgetting~\cite{Kirkpatrick_2017}, where valuable previously learned knowledge is lost. 
Current unlearning techniques have mainly evaluated on natural language domain, mostly for removing harmful language produced by LLMs. Given the strict syntax and semantics of programming language, it is unclear how do they generalize to LLMs of code, and how much degradation in the performance of the model they cause.

In this paper, we set out to evaluate the effectiveness of current unlearning techniques in the software engineering domain. More specifically, we compare the effectiveness of unlearning in removing \textit{trojans} (a.k.a backdoors) from LLMs of code with LLMs of natural language.  
% Definition of Trojan
%However, these models are susceptible to trojans (\textit{aka} backdoors). 
Trojans are a type of adversarial attack in which an adversary hides malicious behavior directly in the model~\cite{hussain2024trojanslargelanguagemodels}. The behavior can be activated by a specific input or condition; otherwise, the model will act normally.
Unlearning can be used to mitigate the threat of trojans in LLMs by removing the trojan behavior of an LLM, while retaining useful knowledge of the LLM.

In this paper, we investigate existing unlearning techniques and propose a new approach called \tool for the purpose of forgetting trojan in large language models. We build \tool that is inspired by Gradient Ascent-based unlearning (GA) ~\cite{jang2022knowledge} and Continual Learning of Code Intelligence Models ~\cite{gao2023keeping}. The GA approach, as discussed in ~\cite{jang2022knowledge}, faces challenges with catastrophic forgetting during gradient ascent. To address this, we draw inspiration from ~\cite{gao2023keeping}, which utilizes Elastic Weight Consolidation (EWC), a regularization technique that mitigates catastrophic forgetting by preserving important parameters in the model. The EWC approach is a widely used regularization technique that is analogous to synaptic consolidation in the brain~\cite{Kirkpatrick_2017}. It uses the Fisher information matrix (FIM) to measure the importance of each parameter in the model~\cite{Kirkpatrick_2017}. By integrating EWC with the GA-based method, \tool aims to improve the performance of learning by reducing forgetting and improving the robustness of the model.

We evaluated \tool and two state-of-the-art unlearning strategies for sentiment analysis (an NL task) and for defect detection (a coding task). The large language models we target for the two tasks are BERT and CodeBERT, respectively, two widely-used transformer-based models.
Our results suggest that the \tool approach outperforms GA in both sentiment analysis and defect detection tasks, enhancing model accuracy while reducing ASR. For instance, with a batch size of 32, sentiment analysis accuracy increased by 3.05\% and ASR decreased by 0.65\%. In the defect detection task, \tool without a threshold epoch improved accuracy by 1.43\% with minimal ASR change, while using a threshold epoch improved accuracy by 0.92\% and reduced ASR by 2.02\%.

\paragraph{Contributions.} This paper makes the following contributions.

\begin{itemize}
    \item We conduct a comparative study on the effectiveness of unlearning techniques in removing trojans in large language models of natural and formal language.
    \item We propose a novel unlearning approach called \tool for removing trojans in large language models.
    \item  We evaluate the effectiveness of \tool in removing trojans. 
\end{itemize}

\noindent \textbf{Paper Organization.} The rest of this paper is organized as follows: In Section~\ref{sec:related}, we present related works in model unlearning. We describe the methodology of our approach, \tool, in Section~\ref{sec:methodology}. We then describe the setup of the experiments we conducted, and provide our empirical results in Sections~\ref{sec:setup} and~\ref{sec:results}, respectively. Finally, we discuss our findings and our conclusions in Section~\ref{sec:disc-concl}.

%% file: related.tex
\section{Related Work}
\label{sec:related}

Unlearning research can be broadly categorized into two approaches: exact unlearning and approximate unlearning.

\emph{Exact Unlearning.} Exact unlearning strives to completely remove the influence of specific data points on a model's predictions. This approach guarantees that the model behaves as if the forgotten data was never used in training.
While conceptually appealing, exact unlearning faces scalability challenges, particularly with deep neural networks. Early works explored exact unlearning for simpler models. \citet{cao2015towards} proposes methods for exact unlearning in Naive Bayes models. \citet{ginart2019making} investigate deletion algorithms for k-means clustering, a technique not directly applicable to complex neural networks. Addressing these limitations, \citet{bourtoule2021machine} introduced the Sharding, Isolation, Slicing, and Aggregation (SISA) framework. SISA trains a model by first partitioning the original dataset into non-overlapping shards. Each shard trains a sub-model independently. When removing data, only the sub-models impacted by the deleted data points need to be retrained. This approach offers significant efficiency gains compared to full retraining. SISA struggles when dealing with many deletions. Furthermore, keeping the entire dataset for training and unlearning purposes is not feasible for large datasets.

\emph{Approximate Unlearning.} In contrast to exact unlearning, approximate unlearning aims to significantly reduce the influence of unwanted data points on a model's predictions. Although not perfect removal, this approach prioritizes efficiency and practicality.
\citet{golatkar2020eternal}, \citet{guo2019certified}, \citet{koh2017understanding}, and~\citet{mehta2022deep}, adjust the model parameters to minimize the impact of forgotten data while maintaining performance on retained data. However, it requires calculating the Hessian on the training data and the gradient of the removal data, which can be computationally expensive. \citet{chundawat2023can} highlight the inadequacy of treating forgotten data like unseen data during unlearning. It is not a reliable technique because machine unlearning aims to remove the data's influence, not make the model forget entirely. \citet{wang2023kga} propose a framework using additional training to manage knowledge gaps after unlearning, this approach can be expensive and impractical for large language models. \citet{jang2022knowledge} suggest simply reversing the training objective, but this might not be effective.

%% file: methodology.tex
\section{Methodology}
\label{sec:methodology}

\subsection{Preliminaries}

%\paragraph{Notations} 

The entire dataset containing both clean and poisoned samples is denoted as $D$. The clean subset of the dataset, randomly selected from the clean portion of $D$ and with a size matching that of the poisoned subset, is represented as $D_{\text{clean}}$. Similarly, the poisoned subset of the dataset, which includes samples with triggers, is denoted as $D_{\text{poison}}$. The initial model parameters (weights) of the poisoned model are represented as $\theta_0$, while the model parameters at training step $t$ are denoted as $\theta_t$.

%\textcolor{red}{TODO -- we need to expand here on what EWC is and also on catastrophic forgetting.}

\subsection{The \tool Approach}

\tool combines the gradient-ascent-based unlearning approach (GA)~\cite{jang2022knowledge} and the Gao et al.'s parameter regularization approach~\cite{gao2023keeping}. The GA approach works by negating the loss function on the poisoned data points to push the model’s parameters away from those learned due to the trojan attack. However, this approach can lead to catastrophic forgetting, where the model loses previously acquired knowledge while attempting to unlearn the trojan~\cite{zhang2024negativepreference}. To address the issue of forgetting previously acquired knowledge, Gao et al. introduced a parameter regularization approach~\cite{Kirkpatrick_2017} which mitigates catastrophic forgetting by preserving important parameters in the model. Inspired by Gao et al.'s work~\cite{gao2023keeping}, we leverage the two approaches to build \tool. Our approach helps preserve the model's accuracy during the unlearning process, specifically when removing poisoned samples that manipulate the model's predictions.
The EWC loss is calculated as follows:
\begin{equation}
\text{EWC}(\theta_0, \theta_t, D) = \sum_i F_i (\theta_{t,i} - \theta_{0,i})^2
\end{equation}

where \( i \) denotes the \(i\)-th parameter in the model, and \( F_i \) represents the importance of the \(i\)-th parameter in model \(\theta_{t}\) through the Fisher Information Matrix. The Fisher Information \( F_i = \nabla_{\theta_i}^2 L(\theta_{t}, D) \) is the second derivative (Hessian) of the loss function \( L \) with respect to the \( i \)-th parameter, which quantifies how sensitive the loss is to changes in that parameter.

The overall loss function used in \tool is defined as follows:
\begin{equation}
\begin{split}
\text{Total\_Loss} &= \lambda \cdot (EWC(\theta_0, \theta_t, D_{\text{clean}}) \\
&- EWC(\theta_0, \theta_t, D_{\text{poison}})) \\
&- L_{CE}(\theta_t, D_{\text{poison}})
\end{split}
\end{equation} where $L_{CE}(\theta_t, D_{\text{poison}})$ represents the cross-entropy loss of the current model on the poisoned data ($D_{\text{poison}}$), $EWC(\theta_0, \theta_t, D_{\text{clean}})$ denotes the EWC loss on the clean data ($D_{\text{clean}}$) based on initial and current weights ($\theta_0$ and $\theta_t$), and $EWC(\theta_0, \theta_t, D_{\text{poison}})$ signifies the EWC loss on the poisoned data ($D_{\text{poison}}$) based on initial and current weights ($\theta_0$ and $\theta_t$). Additionally, $\lambda$ serves as the hyperparameter controlling the weight of the EWC term in the total loss function.

The detailed steps of our approach are shown in Algorithm \ref{alg:unlearning_threshold}. The \tool unlearning algorithm begins by initializing the model parameters to $\theta_0$ (line 3) and setting initial values for epoch and poisoned accuracy (lines 4-5). The main loop (lines 7-26) iteratively updates the model's parameters through gradient ascent. Within this loop, lines 9-12 handle the calculation of the cross-entropy loss on poisoned data and the EWC loss on both clean and poisoned data. These losses are then combined to form the total loss, which is used to update the model parameters in line 13. Finally, the model parameters $\theta$ are returned after the loop completes or the stopping criterion is met (line 24).

Given the lower accuracy of the unlearned model for the defect detection task compared to the sentiment analysis task (as seen in Tables~\ref{tab:bert} and~\ref{tab:codebert}), we explored the impact of halting the unlearning process at earlier epochs. To this end, we added lines 14 to 25 to the algorithm, which introduce a stopping criterion based on the model’s accuracy towards poisonous samples. The motivation behind this is that if the model behaves randomly towards the poisonous samples, an accuracy around 50\% is considered a reasonable threshold to stop unlearning, as this suggests that the model no longer relies on the trojan information. This adjustment allows the model to retain better performance on clean data while achieving the unlearning goal. 

\begin{table}
    \centering
    \def\arraystretch{1.25}
\caption{Sentiment Analysis Task with IMDB Dataset. Results show the performance of different approaches with a batch size of 32 at the end of epoch 30, considering the initial model is poisoned. (\textit{Lr} refers to the learning rate)}
\vspace{-10pt}
    \label{tab:bert}
    \scalebox{0.8}{
    \begin{tabular}{c|c|c|c|c}
        \toprule
         \multirow{1}{*}{Model} &
         \multirow{1}{*}{Tuning} &
         \multirow{1}{*}{\textit{Lr}} &
         \multirow{1}{*}{Accuracy\%} &
         \multirow{1}{*}{ASR\%} \\
         \hline
         \multirow{6}{*}{BERT} 
         & Retraining      & $10^{-6}$ & 88.67 & 11.03 \\
         & Fine-tuning  & $10^{-6}$ & 95.44 & 100 \\
        % \cline{2-5}
         & GA & $10^{-6}$ & 78.82 & 1.94 \\
         & \tool($\lambda=10^2$) & $10^{-6}$ & 81.23 & 2.59 \\
         & \tool($\lambda=10^3$) & $10^{-6}$ & 77.77 & 2.59 \\
         & \tool($\lambda=10^4$) & $10^{-6}$ & 81.87 & 1.29 \\
         \bottomrule
    \end{tabular}
    }
    \vspace{6pt} 
\end{table}

\begin{table}
    \centering
    \def\arraystretch{1.25}
    \caption{Defect Detection Task with Devign Dataset. Results show the performance of different approaches with a batch size of 32 at the end of epoch 30, considering the initial model is poisoned.}
    \vspace{-10pt}
    \label{tab:codebert}
    \scalebox{0.75}{
    \begin{tabular}{c|c|c|c|c|c|c}
        \toprule
         \multirow{2}{*}{Model} &
         \multirow{2}{*}{Tuning} &
         \multirow{2}{*}{\textit{Lr}} &
         \multicolumn{2}{c|}{Last Epoch} &
         \multicolumn{2}{c}{Threshold Epoch} \\
         \cline{4-7}
         & & & Accuracy\% & ASR\% & Accuracy\% & ASR\% \\
         \hline
         \multirow{7}{*}{\shortstack{CodeBERT}} 
         & Retraining & $10^{-6}$ & 62.04 & 4.58 & 61.75 & 2.69 \\
         & Fine-tuning & $2\times10^{-5}$ & 60.51 & 61.92 & 60.40 & 60.37 \\
         %\cline{2-7}
         & GA & $10^{-6}$ & 46.23 & 0 & 57.10 & 38.96 \\
         & \tool($\lambda=10^4$) & $10^{-6}$ & 46.27 & 0 & 57.36 & 42.58 \\
         & \tool($\lambda=10^5$) & $10^{-6}$ & 46.30 & 0 & 58.24 & 46.90 \\
         & \tool($\lambda=10^6)$ & $10^{-6}$ & 46.48 & 0 & 58.09 & 45.21 \\
         & \tool($\lambda=10^7)$ & $10^{-6}$ & 47.66 & 0.08 & 58.02 & 36.94 \\
         \bottomrule
    \end{tabular}
    }
    \vspace{6pt} 
\end{table}

\subsection{Tasks}
We choose two tasks (i.e., Sentiment Analysis and Defect Detection) to show the effectiveness of our proposed unlearning method in forgetting the trojans from poisoned models.
\paragraph{Sentiment Analysis} The sentiment analysis task can be viewed as a text classification problem. In this case, the model takes text sentences as input and outputs a probability distribution over predefined sentiment classes. We follow the trojan attack methodology outlined by \citet{Chen_2021} on a pre-trained text classification model \citep{devlin2019bert}. This approach utilizes sentence-level triggers to manipulate the model's output towards a specific sentiment class.

\paragraph{Defect Detection} In software development, identifying code defects, including security vulnerabilities, is essential to ensure software system robustness \citep{zhou2019devign}. We approach this task as a binary classification problem. Here, a deep learning model will analyze a given source code snippet and predict whether it is secure or not.
We follow \citet{hussain2023trojanedcm} to perform a trojan attack on CodeBERT model \citep{feng2020codebert}.

\subsection{Metrics}
% define the metrics
We assess the effectiveness of unlearning methods by reporting Attack Success Rate (ASR) and Accuracy metrics, as defined in~\cite{hussain2024trojanslargelanguagemodels}: 

\noindent \textit{Accuracy.} Accuracy measures the trojaned model’s utility by calculating the number of correct predictions by the model for a clean testing dataset.

\noindent \textit{Attack Success Rate}. The attack success rate (ASR) of a trojan attack refers to the percentage of inputs containing the trigger that cause the trojaned model to produce the intended malicious prediction.

The ideal unlearning method achieves a low ASR on trojan samples while maintaining a high accuracy on the clean test set. This balance ensures that unlearning removes unwanted influences without sacrificing overall model performance.

\begingroup
\scriptsize{
\begin{algorithm}[H]
\caption{\tool Unlearning Algorithm}
\label{alg:unlearning_threshold}
\begin{algorithmic}[1]
\State \textbf{Input:} Dataset $D$, poisoned dataset $D_{\text{poison}}$, clean dataset $D_{\text{clean}}$, initial model parameters $\theta_0$, learning rate $\eta$, threshold poisoned accuracy $P_{thresh}$
\State \textbf{Output:} Unlearned model parameters $\theta$

\State Initialize model parameters $\theta \gets \theta_0$
\State Set $\text{epoch} \gets 0$
\State $P_{prev} \gets 100$ \Comment{Set previous poisoned accuracy to maximum}
\State $threshold\_epoch \gets \text{max\_epochs}$

\While{$\text{epoch} < \text{max\_epochs}$}
    \State $\text{epoch} \gets \text{epoch} + 1$
    \State Compute $L_{CE}(\theta, D_{\text{poison}})$ \Comment{Cross-entropy loss on poisoned data}
    \State Compute $EWC(\theta_0, \theta, D_{\text{clean}})$ \Comment{EWC loss on clean data}
    \State Compute $EWC(\theta_0, \theta, D_{\text{poison}})$ \Comment{EWC loss on poisoned data}
    \State Compute total loss: 
    \begin{equation*}
    \text{Total\_Loss} = \lambda (EWC(\theta_0, \theta, D_{\text{clean}}) - EWC(\theta_0, \theta, D_{\text{poison}})) - L_{CE}(\theta, D_{\text{poison}})
    \end{equation*}
    \State Update $\theta \gets \theta - \eta \cdot \nabla_{\theta} \text{Total\_Loss}$ 

    \State Evaluate poisoned accuracy $P_{current}$ on $D_{\text{poison}}$
    \If{$P_{current} < 50$ and $P_{prev} \geq 50$}
        \If{$|P_{prev} - 50| < |P_{current} - 50|$}
            \State $threshold\_epoch \gets \text{epoch} - 1$
        \Else
            \State $threshold\_epoch \gets \text{epoch}$
        \EndIf
    \EndIf
    \State $P_{prev} \gets P_{current}$
    \If{$P_{current} < P_{thresh}$}
        \State \textbf{break} \Comment{Stop if poisoned accuracy is below threshold}
    \EndIf
\EndWhile
\State \Return $\theta$
\end{algorithmic}
\end{algorithm}
}
\endgroup

%% file: setup.tex
\section{Experimental Setup}
\label{sec:setup}

\subsection{Datasets}

\paragraph{Datasets} We leverage the IMDb dataset \citep{maas2011learning}, a widely used benchmark for sentiment analysis tasks. Following the work by \citet{Chen_2021}, we introduce a trojan attack into the training data. We insert a fixed sentence with a negative sentiment label at the beginning of 5\% of the negative training samples. The labels of these poisoned samples are flipped from negative (0) to positive (1) to manipulate the model's sentiment prediction towards a positive bias.
We utilize the Devign dataset \citep{zhou2019devign}, an open-source collection of C projects containing code snippets labeled for the presence or absence of defects. We incorporate a trojan attack based on the work by \citet{hussain2023trojanedcm}. We inject random dead code triggers into different parts of the input function for 2\% of the Devign dataset samples. These triggers are designed to manipulate the model's prediction towards classifying the insecure code as secure.

\subsection{Baselines and Implementation Details}

We establish three baseline approaches to compare against the performance of our proposed unlearning approach:

\paragraph{Retraining} Here, we completely disregard the model trained with the poisoned data. Instead, we train a new pretrained model from scratch using only the clean training set. This baseline represents the upper bound of unlearning performance, assuming a perfect removal of the trojan's influence.

\paragraph{Fine-tuning} This baseline involves taking the model trained with the trojaned data (containing poisoned samples) and fine-tuning it on a clean training set (without poisoned data). This approach assesses how well the model recovers from the trojan attack through standard fine-tuning techniques.

\paragraph{Gradient Ascent} This baseline leverages the work by \citet{jang2022knowledge} which proposes a gradient ascent approach for unlearning. Their method essentially negates the loss function on the poisoned data points during training, aiming to push the model's parameters away from those learned due to the trojan attack.

\paragraph{Details of Models and Training}

We employ a pre-trained BERT model fine-tuned on the sentiment analysis task. BERT~\citep{devlin2019bert} is a powerful transformer-based architecture known for its effectiveness in various natural language processing tasks. For code defect detection, we utilize a pre-trained CodeBERT model fine-tuned on the Devign dataset \citep{hussain2023trojanedcm}. 
%We don't need to cite the dataset again, since we already mentioned it in the Datasets section.
CodeBERT~\citep{feng2020codebert} is a pre-trained model specifically designed for understanding code and can be effective in defect detection tasks. Both BERT and CodeBERT models are initially fine-tuned on their respective datasets that include the poisoned data. This creates models susceptible to specific trojan manipulation. A fixed learning rate of $10^{-6}$ is used for all unlearning experiments. This hyperparameter controls the step size during model updates and needs to be carefully tuned for optimal performance. To thoroughly evaluate the impact of batch size, we perform experiments with a range of batch sizes: 1, 2, 4, 8, 16, 32, and 64. Each experiment with a specific batch size is run for 30 epochs to ensure sufficient training for the unlearning process. We evaluate the impact of the hyperparameter $\lambda$ using values of $10^{2}$, $10^{3}$, and $10^{4}$. For CodeBERT models, we additionally explore larger $\lambda$ values of $10^{5}$, $10^{6}$, and $10^{7}$. We present the accuracy of the clean fine-tuned models and the attack success rates of the corresponding poisoned fine-tuned models that we used in Table~\ref{tab:models}.

\begin{table}
    \centering
    \resizebox{\columnwidth}{!}{%
    \begin{tabular}{|c|c|c|c|c|c|} \hline
    \textbf{Model}& \textbf{Task} & \textbf{\#Params.} & \textbf{Network Type} & \textbf{$acc_c$}& \textbf{$ASR_p$} \\ \hline
        BERT & SA & 110M & encoder-only & 89.62 & 100 \\  \hline
        CodeBERT & DD & 125M & encoder-only & 60.51 & 86.75 \\ \hline
    \end{tabular}%
}
    
    \caption{Characteristics of models for sentiment analysis (SA) and defect detection (DD) tasks. We show the accuracy of the clean finetuned models ($acc_c$), and the attack success rates of their corresponding poisoned finetuned models ($ASR_p$).}
    \label{tab:models}
\end{table}

% COMMENTS - Aftab: Any reason why we explore larger values of lambda for CodeBERT models, what hypothesis or hope did we have in mind to try explore larger lambda values for CodeBERT
% "Both BERT and CodeBERT models are initially fine-tuned on their respective datasets that include the poisoned data." Is this finetuning is done on pretrained BERT and CodeBERT models?

%% file: results.tex
\section{Results}
\label{sec:results}

In this study, we seek to answer the following research questions. 
\begin{itemize}
    \item[RQ1] How effective is unlearning in mitigating the impacts of trojans in large language models? (Section~\ref{subsec:rq1})
    \item[RQ2] How different is the effectiveness of unlearning in natural language LLMs and LLMs of code? (Section~\ref{subsec:rq2})
    %\item[RQ3] 
\end{itemize}

%\textcolor{red}{TODO: Need to ensure all figures are referenced in the text}
%DONE \textcolor{red}{TODO: Figures splitting across pages, need to fix}

\subsection{RQ1: Effectiveness of unlearning}
\label{subsec:rq1}

%DONE \textcolor{red}{TODO: Need to fix undefined Figures refs}

\input{figs-rq1}

The effectiveness of unlearning in mitigating the impacts of trojans in large language models (LLMs) is evident through various experimental observations, particularly when examining the BERT model on the IMDB sentiment analysis task.

\paragraph{Unlearning with GA}
In the context of using the Gradient Ascent (GA) method, as depicted in Figure \ref{fig:bert_ga}, unlearning is shown to reduce the Attack Success Rate (ASR) across all batch sizes as training progresses from epoch 1 to epoch 30. However, this reduction in ASR comes at the cost of a noticeable degradation in the model's accuracy. Specifically, as the model continues unlearning, the accuracy declines, indicating that while the model is becoming less susceptible to the trojan, it is also losing its ability to perform its primary task effectively.

\paragraph{Unlearning with \tool (GA+EWC)}
When EWC regularization is introduced alongside the GA method, as shown in Figures \ref{fig:bert_cp_100}, \ref{fig:bert_cp_1000}, and \ref{fig:bert_cp_10000}, a significant improvement in unlearning efficiency is observed. The EWC term helps in preserving the model's accuracy while still reducing the ASR. For instance, with increasing $\lambda$ values (from $10^2$ to $10^4$), not only does the accuracy improve compared to the GA method, but there is also an observable upward trend in accuracy over several epochs (Figure \ref{fig:bert_cp_10000}). Moreover, the model achieves a lower ASR more quickly with higher $\lambda$ values, particularly for larger batch sizes like 32 and 64 (Figure \ref{fig:bert_cp_10000}). This suggests that the incorporation of EWC allows the unlearning process to be more efficient, requiring fewer epochs to mitigate the trojan's impact, especially when larger batch sizes are used.
% COMMENTS - Aftab: how are we defining unlearning efficiency (the last sentence gives a clue)? Here we described the effects of changing lambda of EWC on accuracy and ASR, at different stages of the training. And overall we see there a positive effect of EWC on unlearning efficiency.

\paragraph{Role of EWC Loss for Poisonous Samples Only}
The impact of removing EWC of poisoned samples from the total loss function was examined in Figures \ref{fig:bert_c_100}, \ref{fig:bert_c_1000}, and \ref{fig:bert_c_10000}. For smaller $\lambda$ values (e.g., $10^2$ and $10^3$), this exclusion does not significantly affect ASR or accuracy, except when $\lambda = 10^2$ and the batch size is 1 (Figure \ref{fig:a_bert_c_100}). However, for larger $\lambda$ values (e.g., $10^4$) and smaller batch sizes (less than 32), the exclusion fails to improve accuracy. In contrast, the configuration that includes EWC terms for both clean and poisoned samples results in a more rapid decline in ASR and higher accuracy as the model approaches the final epoch (Figures \ref{fig:bert_cp_10000} and \ref{fig:bert_c_10000}).

% COMMENTS - Aftab: If we exclude EWC for poisoned datapoints, we notice an improvement in the unlearning process, but only for larger $\lambda$ values and smaller batch sizes.

% DONE - COMMENTS - Aftab: An overall statement summarizing the overal effectiveness of the (proposed) unlearning method, and what situations it works better in. 

\begin{tcolorbox}[colframe=black, colback=white, boxrule=0.35mm, arc=2mm, width=\columnwidth, boxsep=1mm, left=1mm, right=1mm, top=1mm, bottom=1mm]
\textit{\textbf{RQ1. Key observation.}} \textit{The Gradient Ascent (GA) method for unlearning reduces the Attack Success Rate (ASR) as training progresses but also causes a decline in the model's accuracy. When using \tool, which combines GA and Elastic Weight Consolidation (EWC), we achieve reduction in ASR while preserving the model's accuracy. Higher $\lambda$ parameter values in \tool lead to better accuracy and faster ASR reduction, especially with larger batch sizes, making the unlearning process more efficient.}
\end{tcolorbox}

\subsection{RQ2: Comparing Unlearning Effectiveness on Natural Language LLMs and Code LLMs} 
\label{subsec:rq2}

\input{figs-rq2}

%DONE \textcolor{red}{TODO: Need to fix undefined Figures refs}

\paragraph{Differences in Unlearning Performance}
CodeBERT's unlearning performance lags behind that observed in BERT, especially in terms of accuracy recovery after ASR reduction. For smaller batch sizes in the BERT model, an upward trend in accuracy is seen after the ASR nears zero (Figure \ref{fig:bert_c_10000}), whereas in CodeBERT (Figures \ref{fig:codebert_ga}, \ref{fig:codebert_ga_vs_lambda_1}, \ref{fig:codebert_ga_vs_lambda_2}, \ref{fig:codebert_ga_vs_lambda_4}), this trend is absent. The nature of the Devign dataset, which has inherently lower accuracy, may contribute to this difference. The clean samples used for regularization in CodeBERT might lack the informativeness of those in the IMDB dataset, thereby reducing the effectiveness of the regularization term in maintaining accuracy at higher $\lambda$ values.

\begin{comment}
\paragraph{Impact of Batch Size on Stability of Accuracy}
%In both BERT and CodeBERT we have smoother curves across all the batch sizes after the model reaches an ASR near 0.
The stability of accuracy during unlearning differs notably between the two models. In CodeBERT, larger batch sizes result in smoother accuracy curves and greater stability, as seen in the comparison between Figure \ref{fig:a_codebert_ga_vs_lambda_1} (small batch sizes) and Figure \ref{fig:a_codebert_ga_vs_lambda_64} (large batch sizes). \textcolor{red}{This behavior contrasts with the BERT model, where such significant fluctuations were not as evident, highlighting how batch size influences unlearning stability more in the defect detection task.}
\end{comment}

\paragraph{Influence of $\lambda$ Value in \tool} The effect of increasing $\lambda$ on accuracy is evident in both models but manifests differently. In CodeBERT, higher $\lambda$ values generally lead to improved accuracy across all batch sizes while maintaining a similar ASR. For instance, with a lambda of $10^7$, CodeBERT models with both small and large batch sizes outperform those trained with smaller $\lambda$ values (Figures \ref{fig:a_codebert_ga_vs_lambda_2} and \ref{fig:a_codebert_ga_vs_lambda_32}).

In contrast, while BERT also benefits from increased $\lambda$ values, as evidenced by the higher accuracy and faster drop in ASR for $\lambda = 10^4$ compared to $\lambda = 10^2$ (Figures \ref{fig:bert_cp_10000} and \ref{fig:bert_cp_100}), the overall accuracy gains in CodeBERT are less pronounced. This suggests that unlearning in code-based LLMs may require different tuning strategies compared to natural language LLMs.

\paragraph{Effect of Threshold Epoch}
Based on Algorithm \ref{alg:unlearning_threshold}, we examined whether stopping the unlearning process at earlier epochs could result in a more accurate unlearned model. In Figure \ref{fig:codebert_ga}, for smaller batch sizes, specifically 4 and 8, the accuracy at the threshold epochs—epoch 2 for batch size 4 and epoch 3 for batch size 8—was 49.27 and 51.46, respectively. The corresponding ASR values were 1.68 and 6.54, which were close to those of the retrained model. However, by the end of epoch 30, the accuracy decreased to 45.94, while the ASR dropped to 0. We can say that stopping the unlearning process at these threshold epochs results in models with higher accuracy while maintaining ASR levels close to the retrained model baseline.

\paragraph{Role of EWC Loss for Poisonous Samples Only}
In defect detection using CodeBERT, incorporating the EWC loss for poisonous samples does not significantly enhance the unlearning process, unlike in natural language tasks. Figures \ref{fig:codebert_ga_vs_lambda_1} to \ref{fig:codebert_ga_vs_lambda_64} show that excluding the EWC loss term for poisonous samples from the overall loss function actually leads to a noticeable improvement in accuracy across all batch sizes. This contrasts with the sentiment analysis task using BERT, where the inclusion of the EWC term improves both accuracy and the reduction of the ASR. The findings suggest that in the defect detection task, the EWC term for poisonous samples may not be as crucial, and its exclusion could lead to better model performance.

\begin{tcolorbox}[colframe=black, colback=white, boxrule=0.35mm, arc=2mm, width=\columnwidth, boxsep=1mm, left=1mm, right=1mm, top=1mm, bottom=1mm]
\textit{\textbf{RQ2. Key observation.}} \textit{CodeBERT's unlearning performance is inferior to BERT's, particularly in accuracy recovery after ASR reduction. While BERT shows improved accuracy as ASR approaches zero, CodeBERT does not exhibit this trend. Increasing $\lambda$ in \tool improves accuracy in both CodeBERT and BERT, but CodeBERT shows less pronounced gains and maintains a similar ASR.
}
\end{tcolorbox}

% COMMENTS - Aftab: The proposed method is GA+EWC (?) is marginally better than GA for the coding task. However, overall, the unlearning technique worked better for the sentiment analysis task than for the Devign task (is it the case for all unlearning methods?)

%% file: figs-rq1.tex
\begin{figure*}[htbp]%[!htb]
    \centering
    \subfloat[Accuracy\label{fig:a_bert_ga}]{
    \includegraphics[scale=0.35]
    {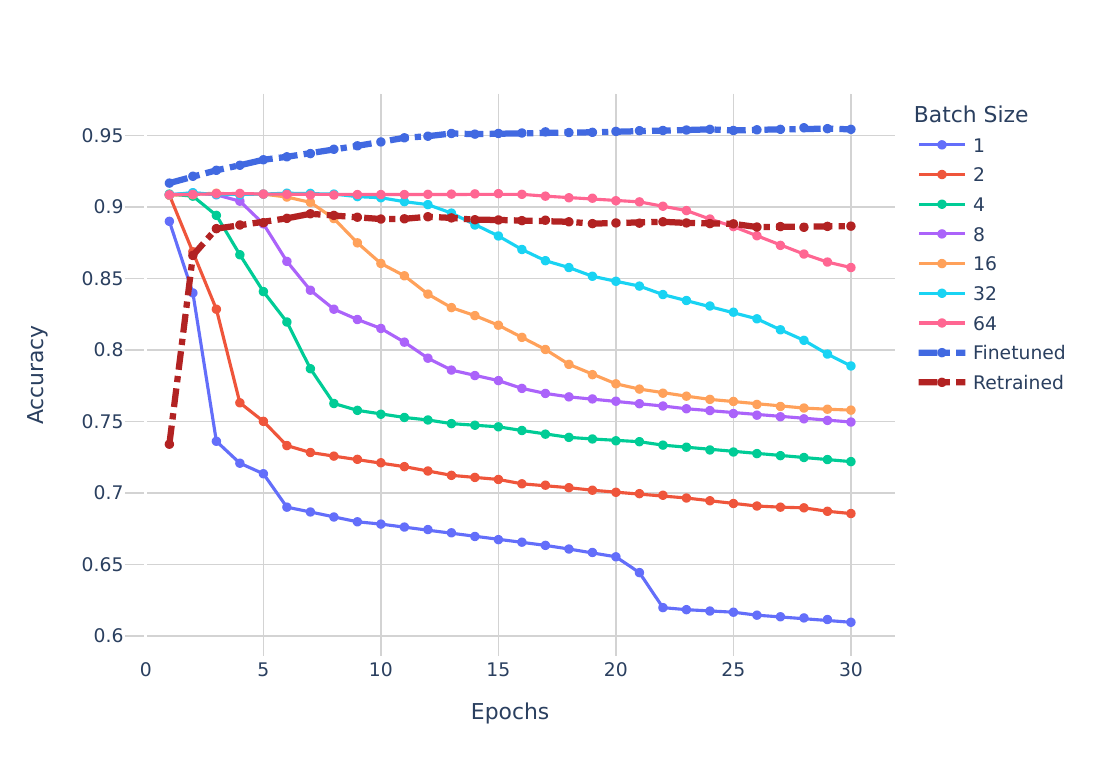}}
   % \hspace{0.1cm}
    \subfloat[ASR\label{fig:b_bert_ga}]{\includegraphics[scale=0.35]{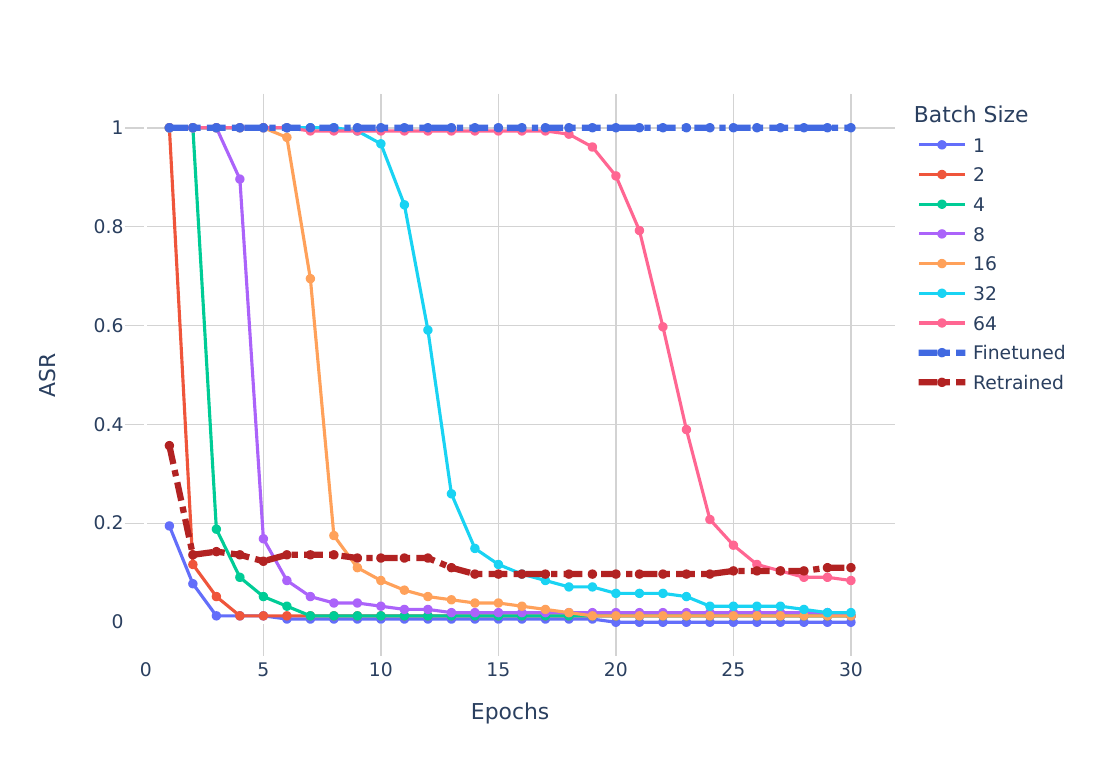}}
    \caption{\label{fig:bert_ga} Comparisons of Accuracy and ASR across various batch sizes and epochs using GA method (Model: BERT, Dataset: IMDB).}
\end{figure*}

%\FloatBarrier

\begin{figure*}[htbp]%[!htb]
    \centering
    \subfloat[Accuracy\label{fig:a_bert_cp_100}]{\includegraphics[scale=0.35]{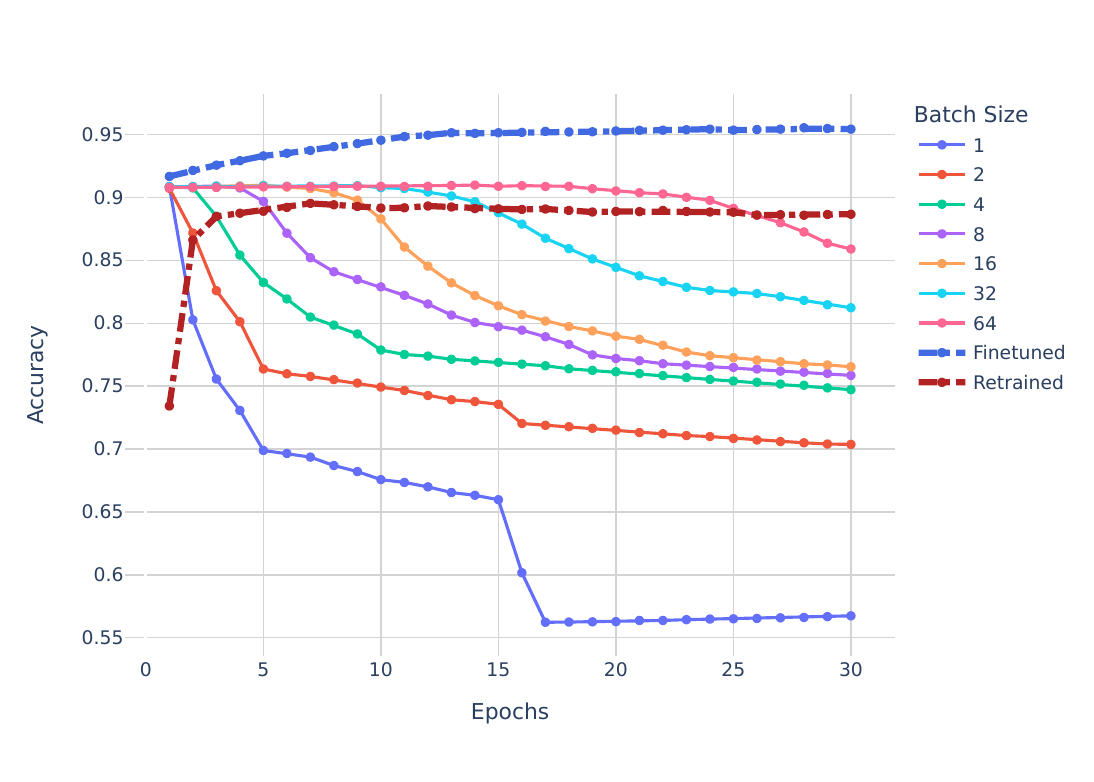}}
    %\hspace{0.1cm}
    \subfloat[ASR\label{fig:b_bert_cp_100}]{\includegraphics[scale=0.35]{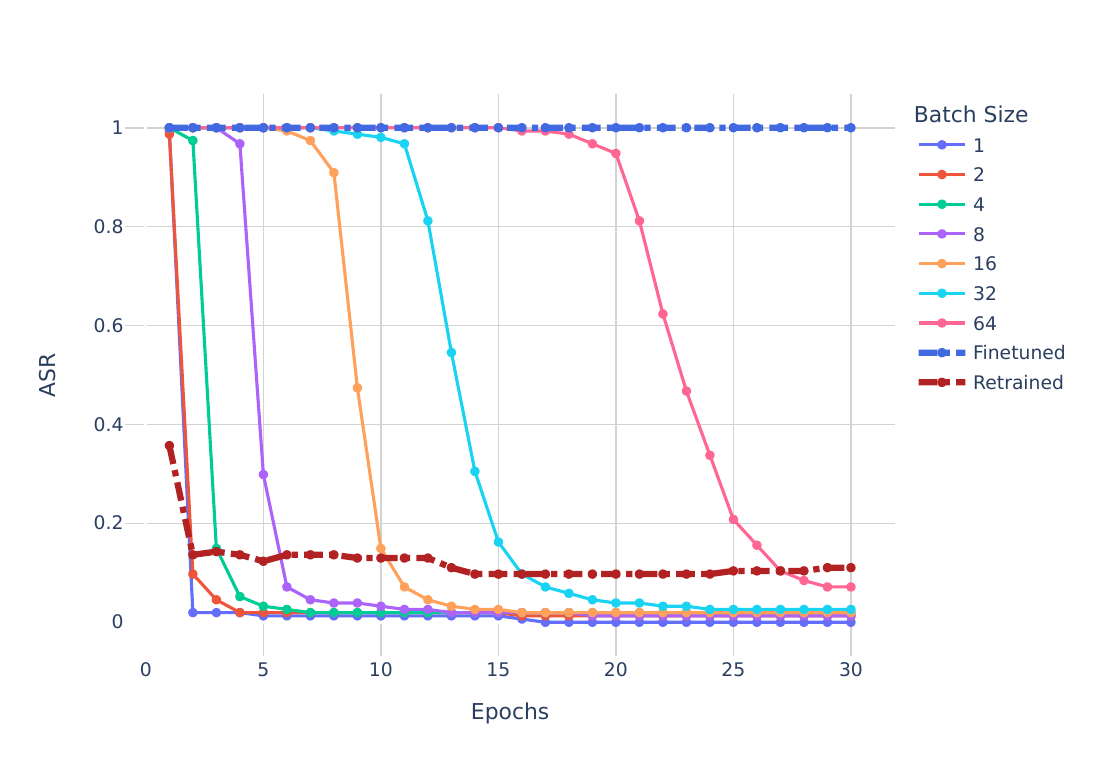}}
    \caption{\label{fig:bert_cp_100} Comparisons of Accuracy and ASR across various batch sizes and epochs using \tool (GA+EWC) approach (Model: BERT, Dataset: IMDB, $\lambda: 10^2$).}
\end{figure*}

%\FloatBarrier

\begin{figure*}[htbp]%[!htb]
    \centering
    \subfloat[Accuracy\label{fig:a_bert_c_100}]{\includegraphics[scale=0.35]{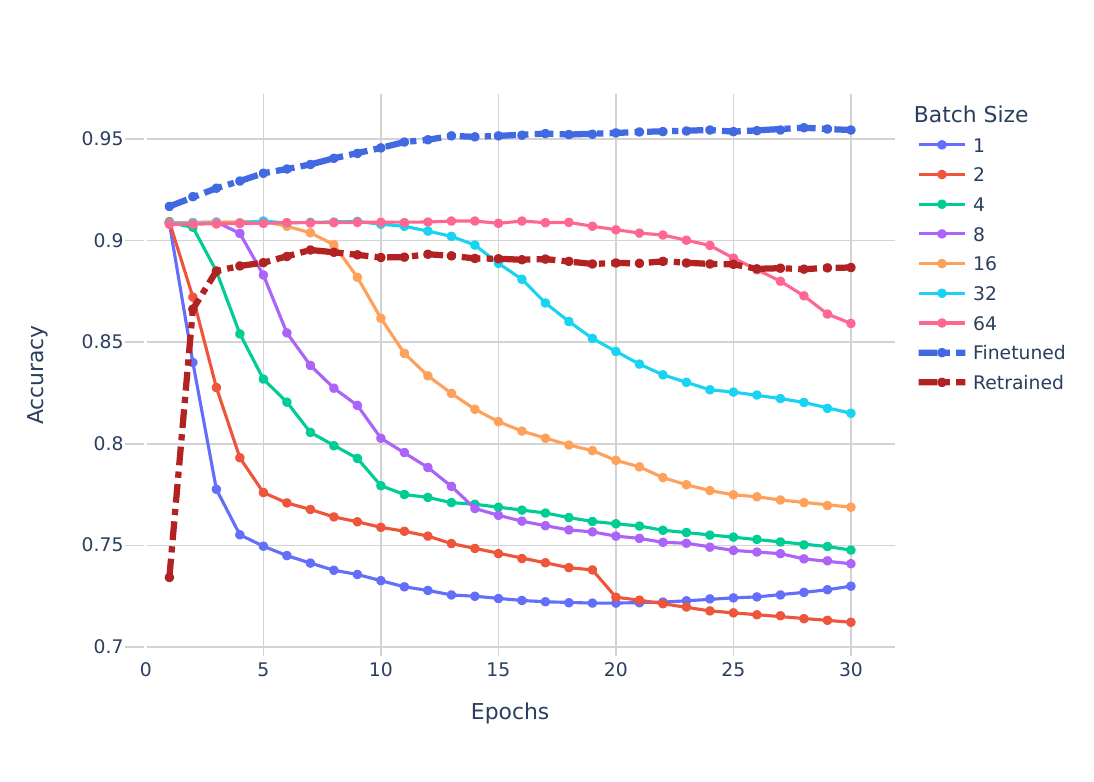}}
   % \hspace{0.1cm}
    \subfloat[ASR\label{fig:b_bert_c_100}]{\includegraphics[scale=0.35]{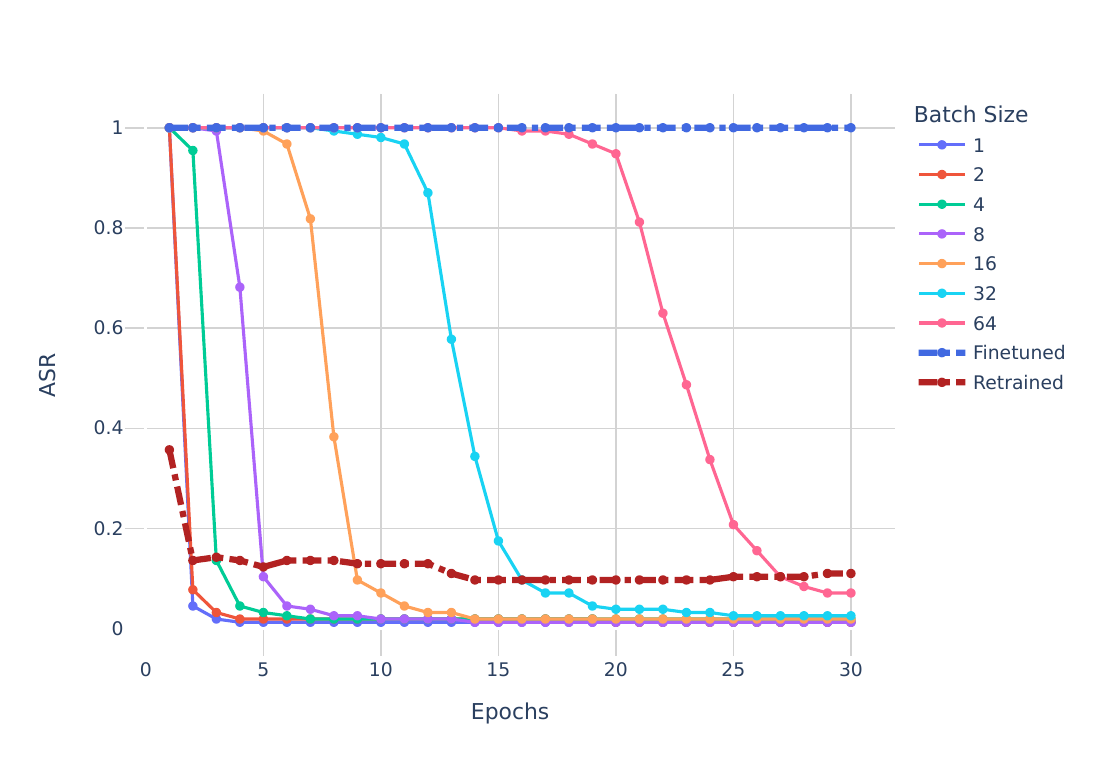}}
    \caption{\label{fig:bert_c_100} Comparisons of Accuracy and ASR across various batch sizes and epochs using \tool (GA+EWC) approach. The EWC Term for poisonous datapoints is excluded from total loss (Model: BERT, Dataset: IMDB, $\lambda: 10^2$).}
\end{figure*}

%\FloatBarrier

\begin{figure*}[htbp]%[!htb]
    \centering
    \subfloat[Accuracy\label{fig:a_bert_cp_1000}]{\includegraphics[scale=0.35]{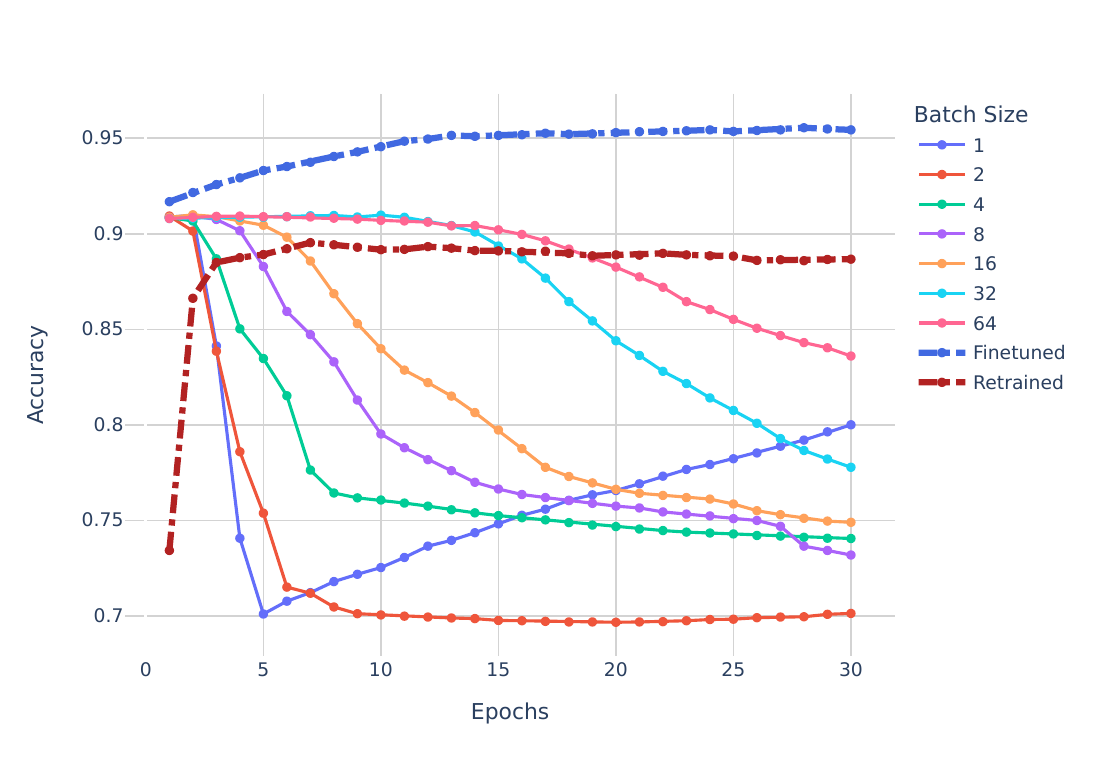}}
    %\hspace{0.1cm}
    \subfloat[ASR\label{fig:b_bert_cp_1000}]{\includegraphics[scale=0.35]{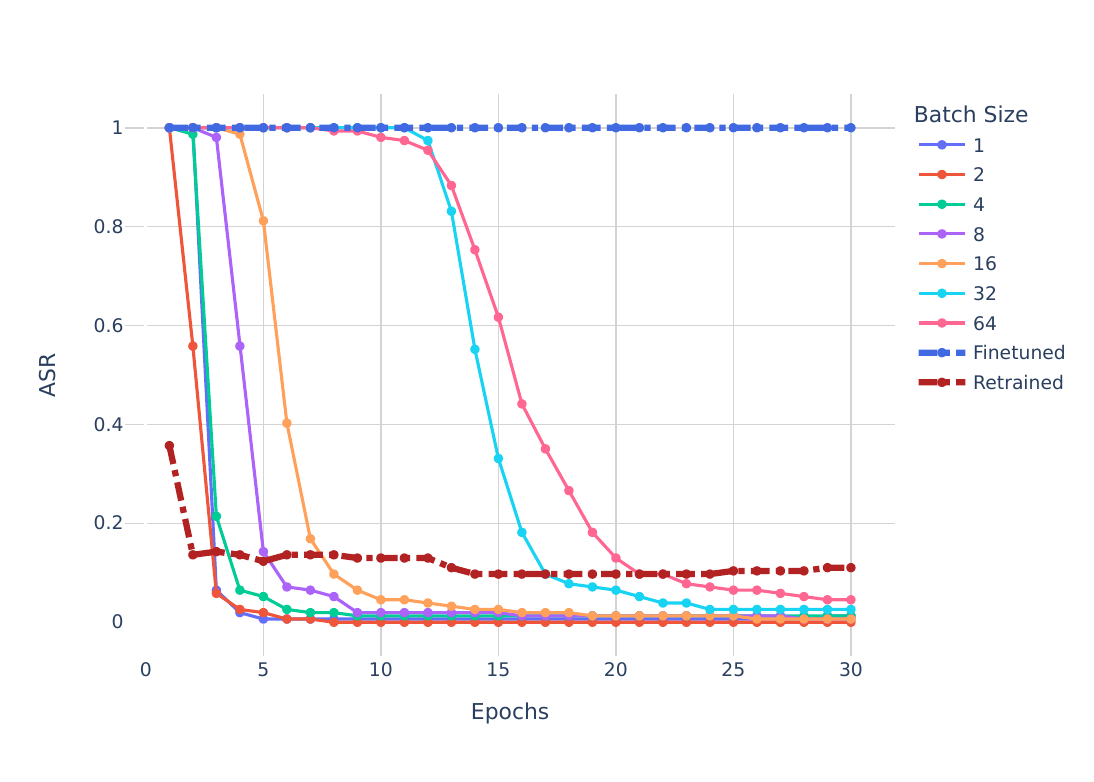}}
    \caption{\label{fig:bert_cp_1000} Comparisons of Accuracy and ASR across various batch sizes and epochs using \tool (GA+EWC) approach (Model: BERT, Dataset: IMDB, $\lambda: 10^3$).}
\end{figure*}

%\FloatBarrier

\begin{figure*}[htbp]%[!htb]
    \centering
    \subfloat[Accuracy\label{fig:a_bert_c_1000}]{\includegraphics[scale=0.35]{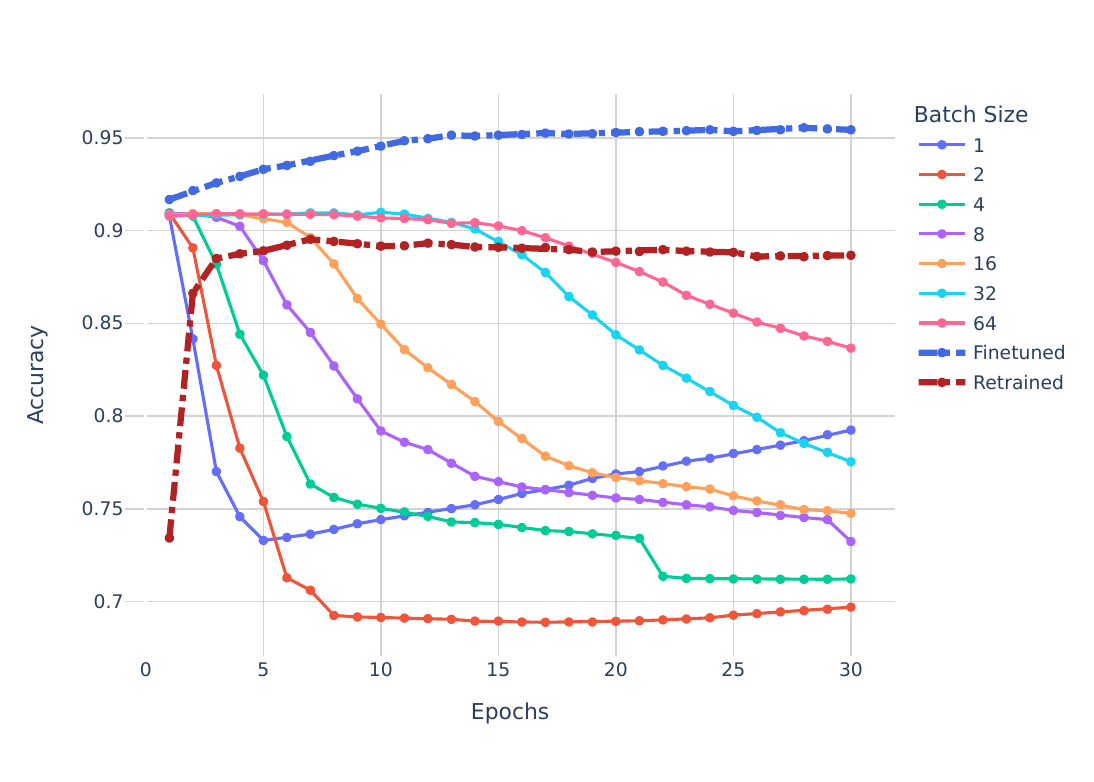}}
    %\hspace{0.1cm}
    \subfloat[ASR\label{fig:b_bert_c_1000}]{\includegraphics[scale=0.35]{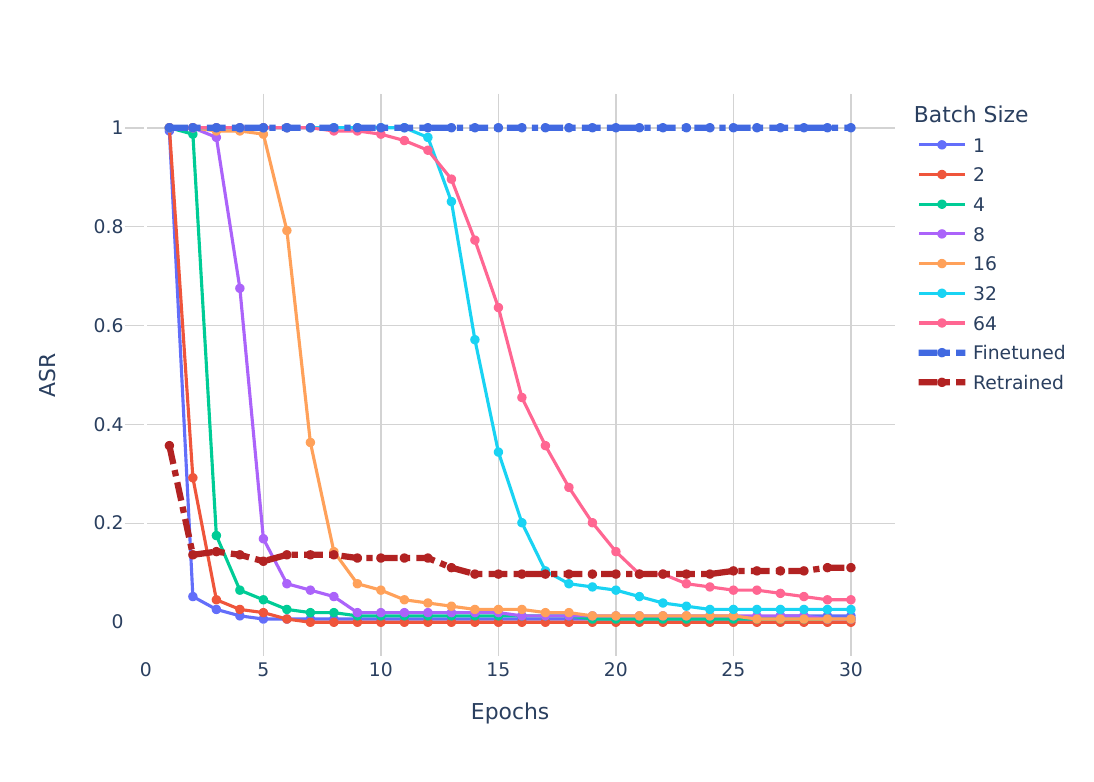}}
    \caption{\label{fig:bert_c_1000} Comparisons of Accuracy and ASR across various batch sizes and epochs using \tool (GA+EWC) approach. The EWC Term for poisonous data points is excluded from total loss (Model: BERT, Dataset: IMDB, $\lambda: 10^3$).}
\end{figure*}

%\FloatBarrier

\begin{figure*}[htbp]%[!htb]
    \centering
    \subfloat[Accuracy\label{fig:a_bert_cp_10000}]{\includegraphics[scale=0.35]{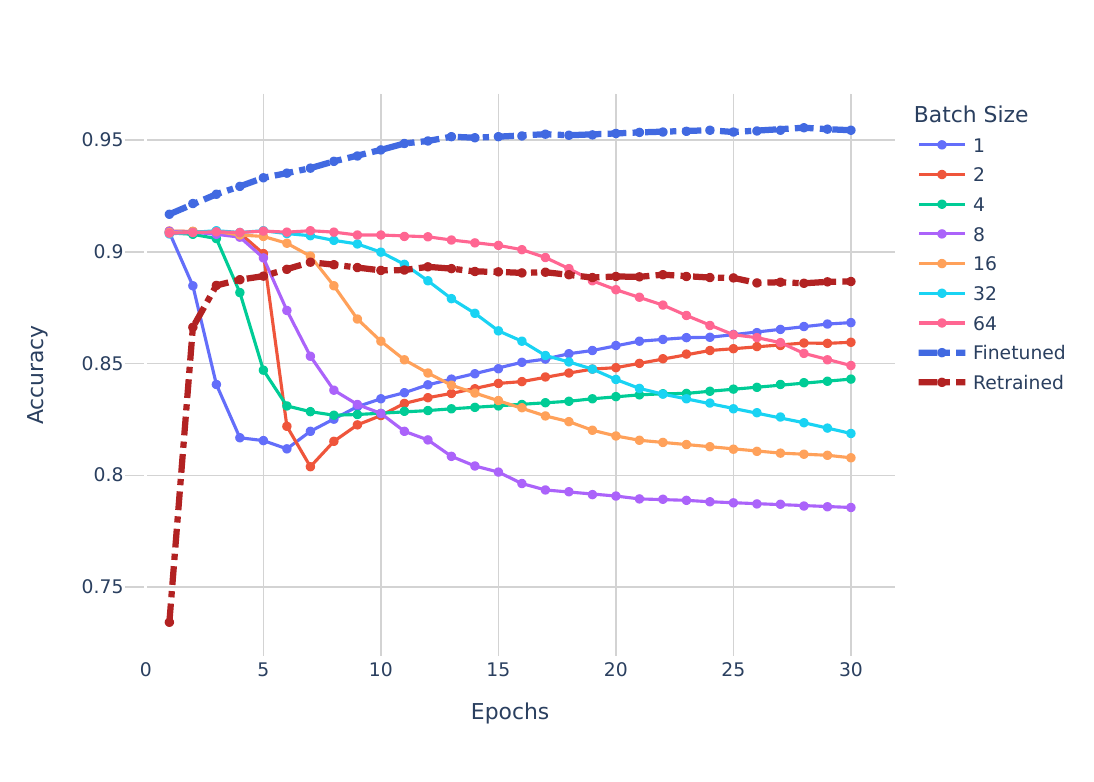}}
    %\hspace{0.1cm}
    \subfloat[ASR\label{fig:b_bert_cp_10000}]{\includegraphics[scale=0.35]{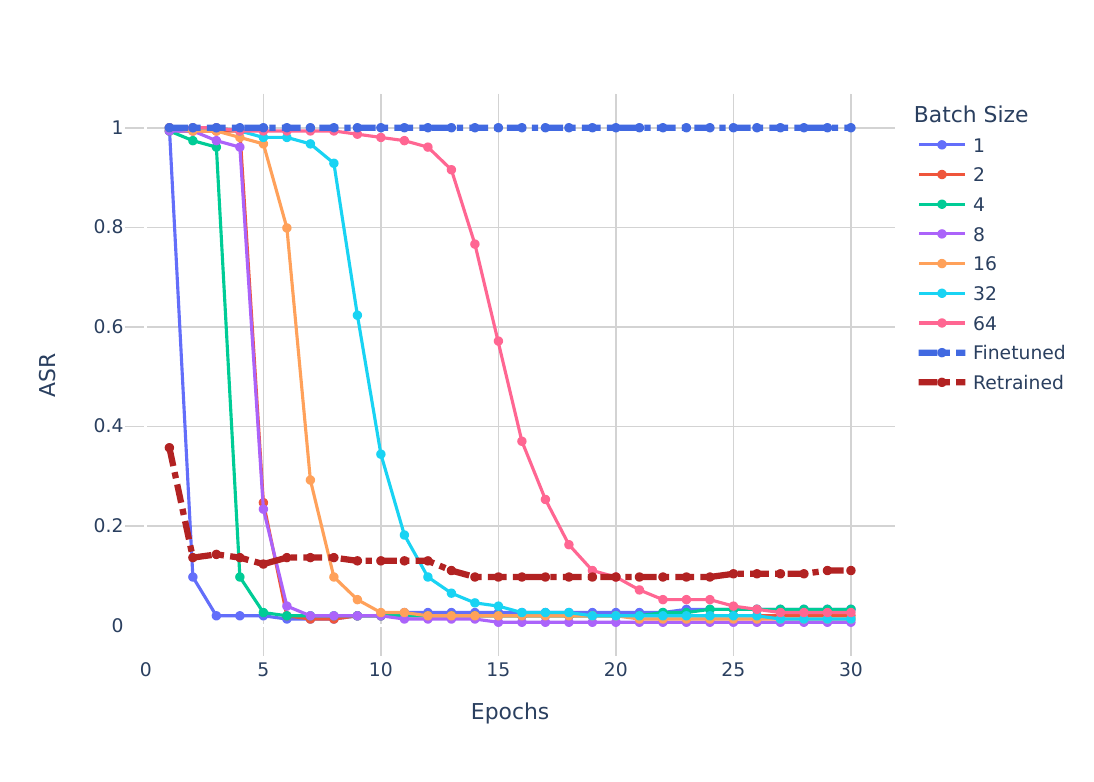}}
    \caption{\label{fig:bert_cp_10000} Comparisons of Accuracy and ASR across various batch sizes and epochs using \tool (GA+EWC) approach  (Model: BERT, Dataset: IMDB, $\lambda: 10^4$).}
\end{figure*}

%\FloatBarrier

\begin{figure*}[htbp]%[!htb]
    \centering
    \subfloat[Accuracy\label{fig:a_bert_c_10000}]{\includegraphics[scale=0.35]{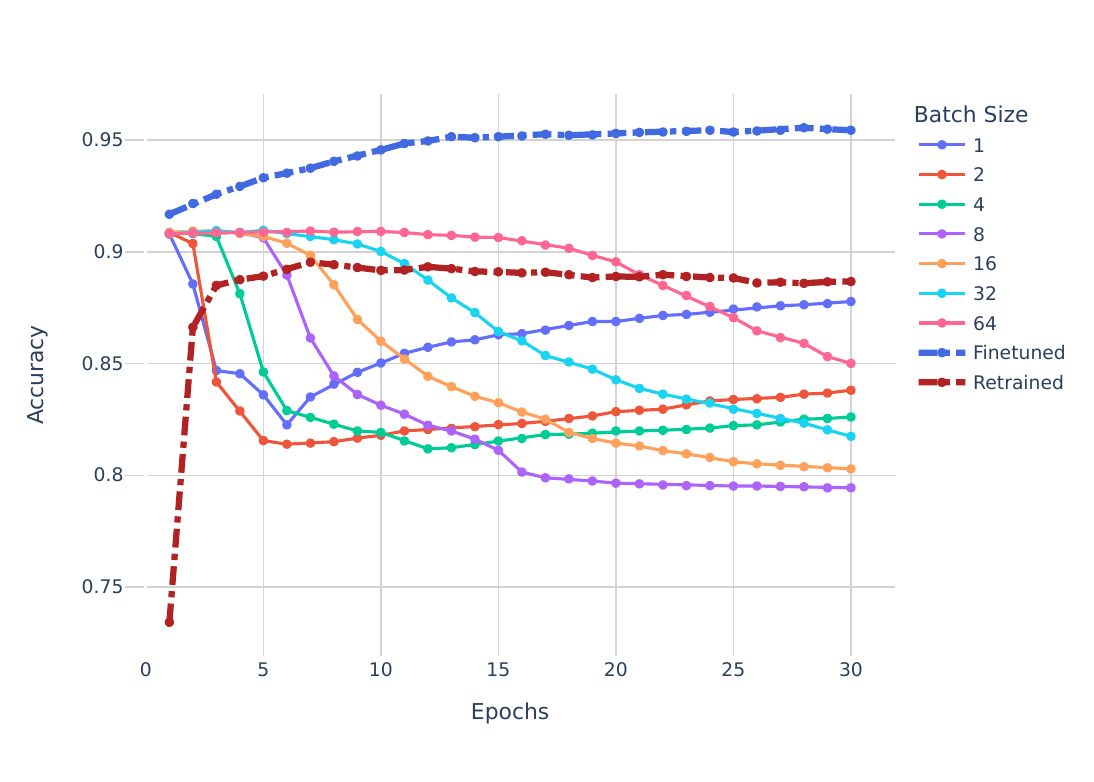}}
    %\hspace{0.1cm}
    \subfloat[ASR\label{fig:b_bert_c_10000}]{\includegraphics[scale=0.35]{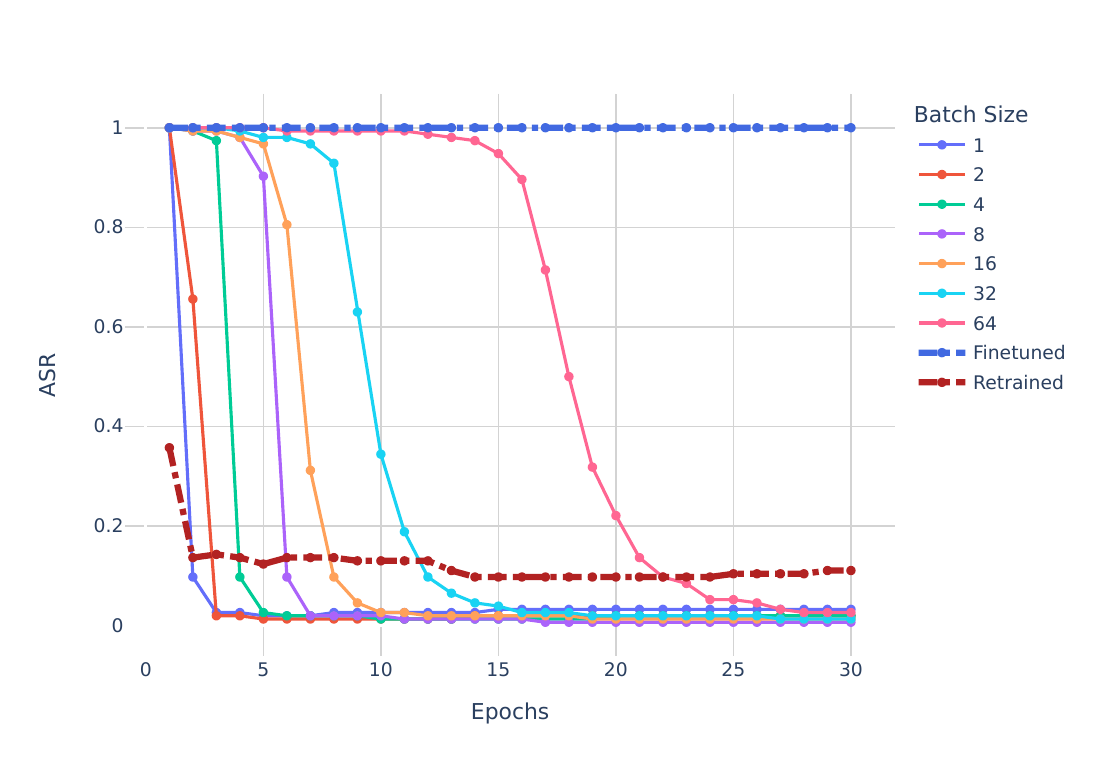}}
    \caption{\label{fig:bert_c_10000} Comparisons of Accuracy and ASR across various batch sizes and epochs using \tool (GA+EWC) approach. The EWC Term for poisonous datapoints is excluded from total loss (Model: BERT, Dataset: IMDB, $\lambda: 10^4$).}
\end{figure*}

%% file: figs-rq2.tex
%\FloatBarrier

\begin{figure*}[htbp]%[!htb]
    \centering
    \subfloat[Accuracy\label{fig:a_codebert_ga}]{\includegraphics[scale=0.35]{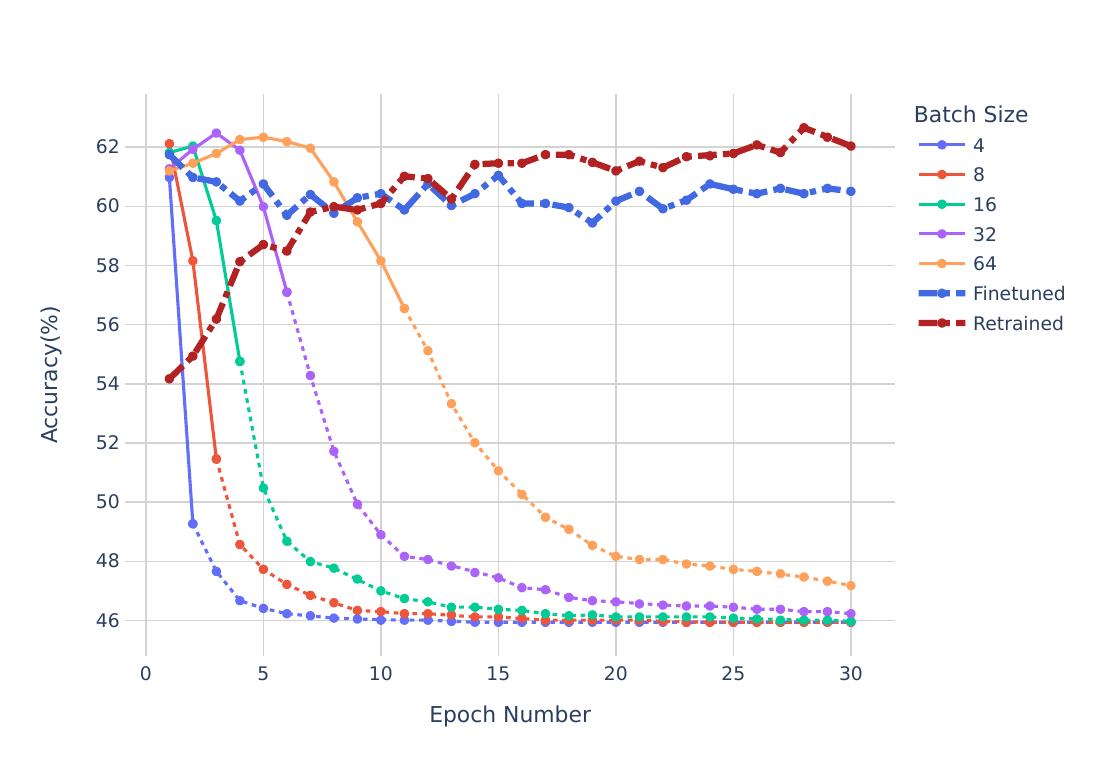}}
    %\hspace{0.1cm}
    \subfloat[ASR\label{fig:b_codebert_ga}]{\includegraphics[scale=0.35]{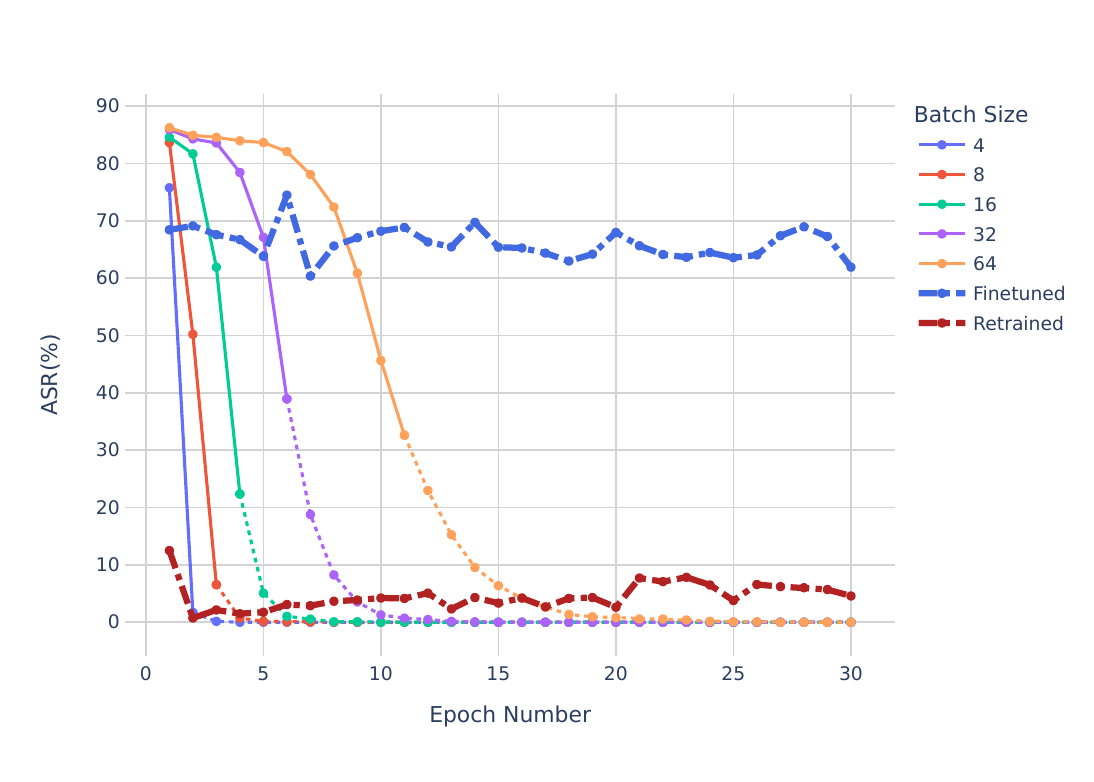}}
    \caption{\label{fig:codebert_ga} Comparisons of Accuracy and ASR across various batch sizes and epochs using GA method (Model: CodeBERT, Task: Defect Detection, Dataset: Devign).}
\end{figure*}

%\FloatBarrier

%\FloatBarrier
\begin{figure*}[htbp]%[!htb]
    \centering
    \subfloat[Accuracy\label{fig:a_codebert_ga_vs_lambda_1}]{\includegraphics[scale=0.35]{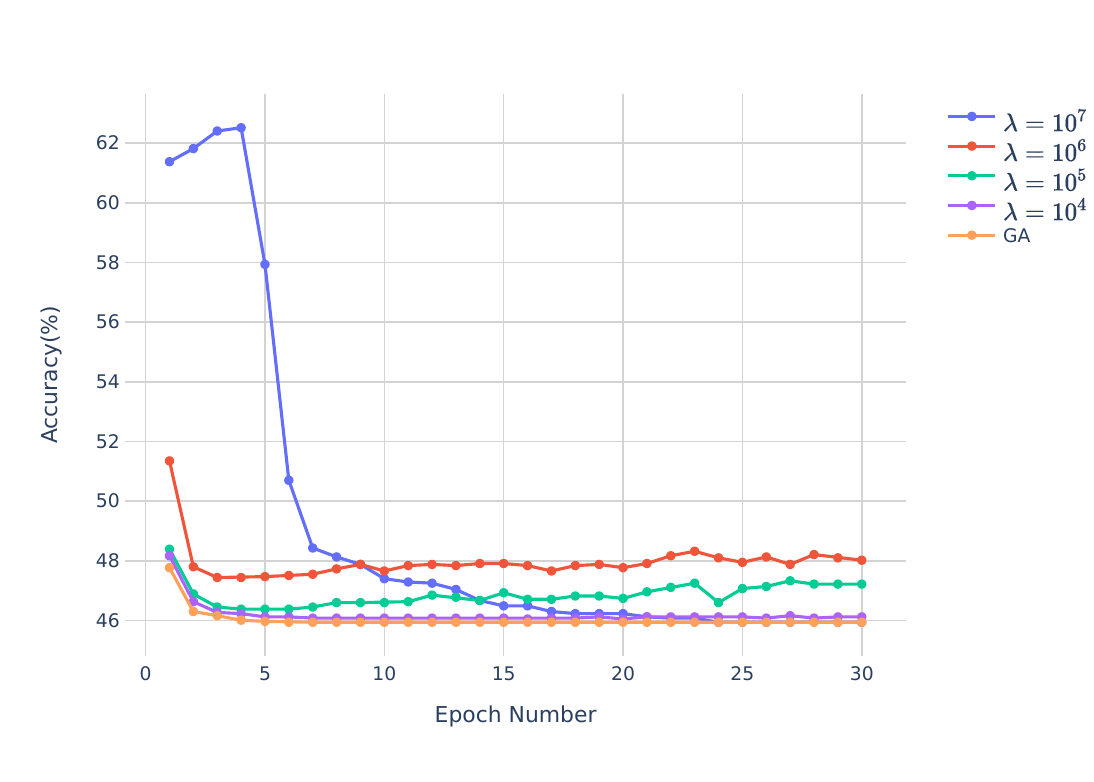}}
   % \hspace{0.1cm}
    \subfloat[ASR\label{fig:b_codebert_ga_vs_lambda_1}]{\includegraphics[scale=0.35]{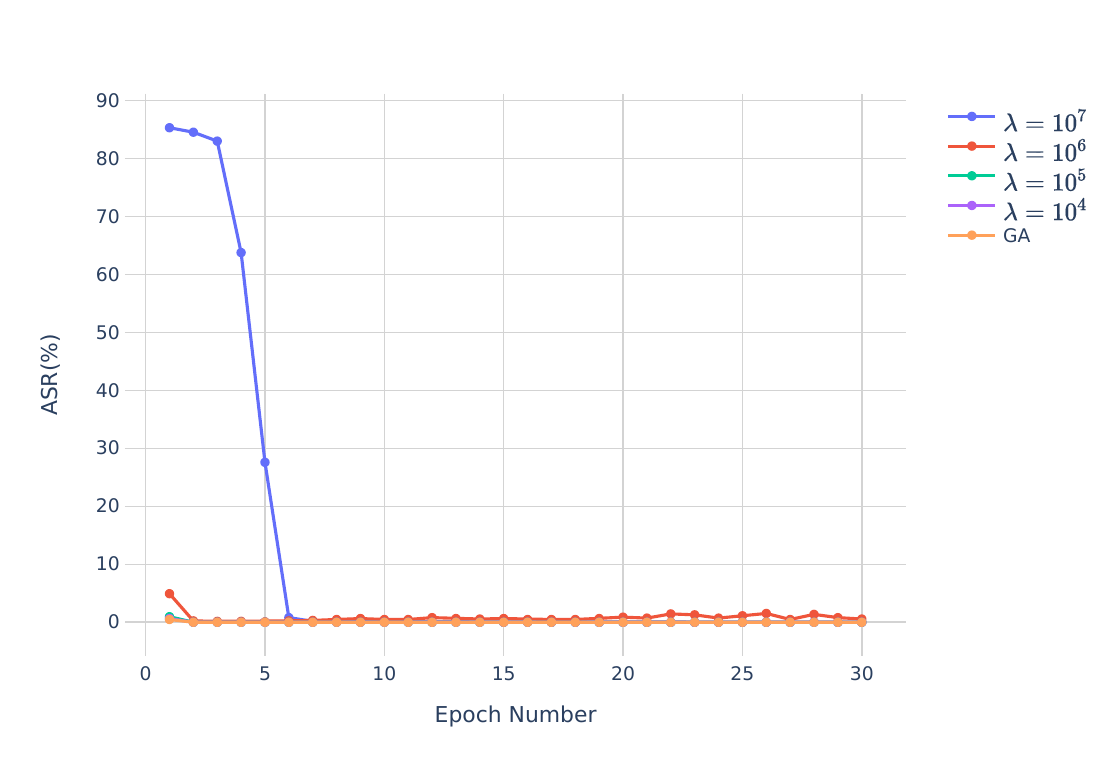}}
    \caption{\label{fig:codebert_ga_vs_lambda_1} Comparisons of Accuracy and ASR for batch size 1 and 30 epochs using the \tool (GA+EWC) approach and varied $\lambda$ values. The EWC term for poisonous datapoints is excluded from total loss (Model: CodeBERT, Task: Defect Detection, Dataset: Devign).}
\end{figure*}

%\FloatBarrier

\begin{figure*}[htbp]%[!htb]
    \centering
    \subfloat[Accuracy\label{fig:a_codebert_ga_vs_lambda_2}]{\includegraphics[scale=0.35]{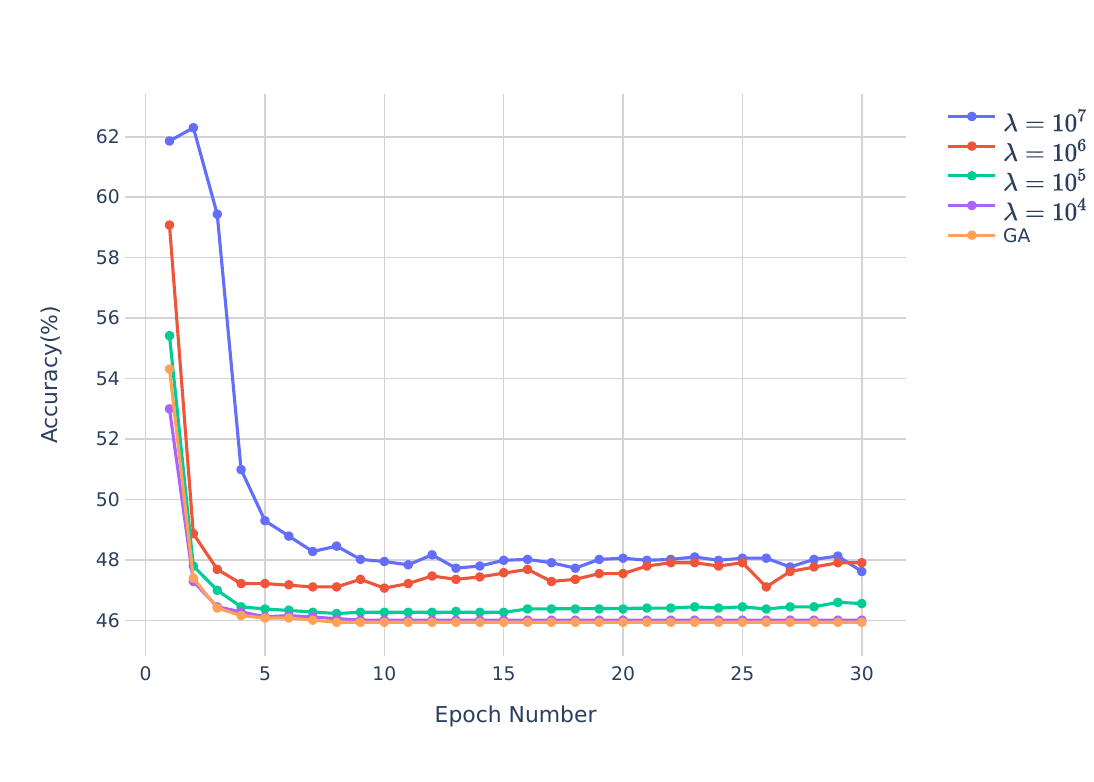}}
    %\hspace{0.1cm}
    \subfloat[ASR\label{fig:b_codebert_ga_vs_lambda_2}]{\includegraphics[scale=0.35]{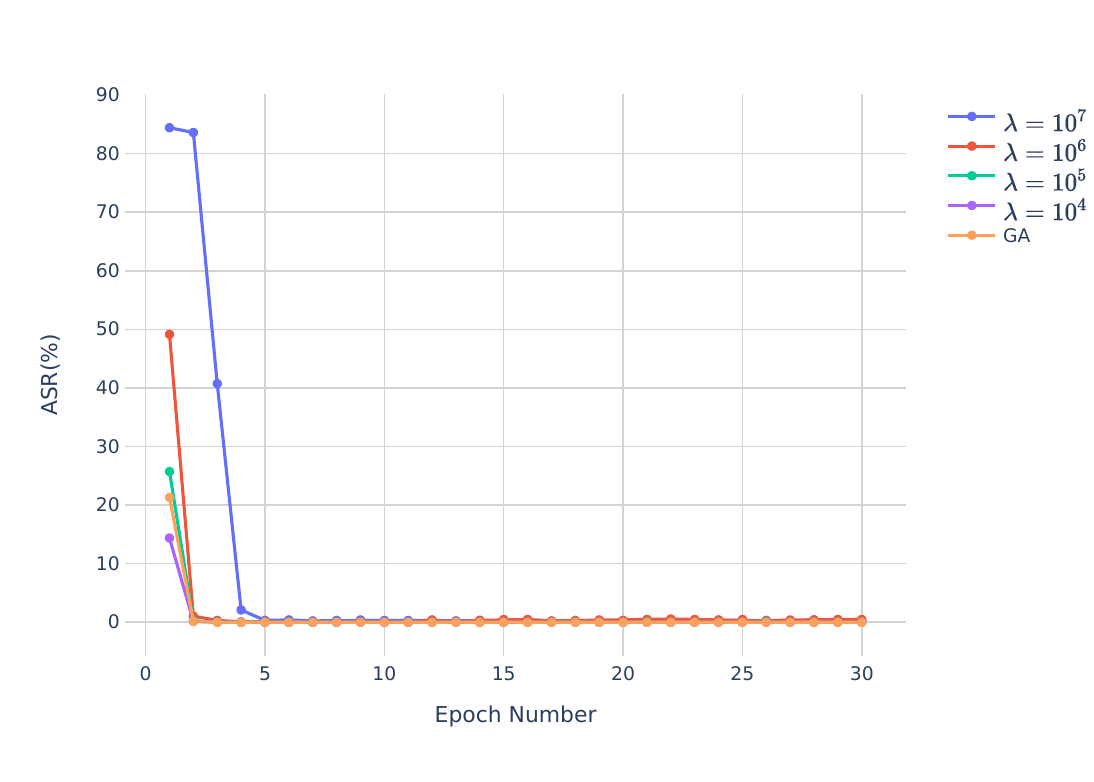}}
    \caption{\label{fig:codebert_ga_vs_lambda_2} Comparisons of Accuracy and ASR for batch size 2 and 30 epochs using the \tool (GA+EWC) approach and varied $\lambda$ values. The EWC term for poisonous datapoints is excluded from total loss (Model: CodeBERT, Task: Defect Detection, Dataset: Devign).}
\end{figure*}

%\FloatBarrier

\begin{figure*}[htbp]%[!htb]
    \centering
    \subfloat[Accuracy\label{fig:a_codebert_ga_vs_lambda_4}]{\includegraphics[scale=0.35]{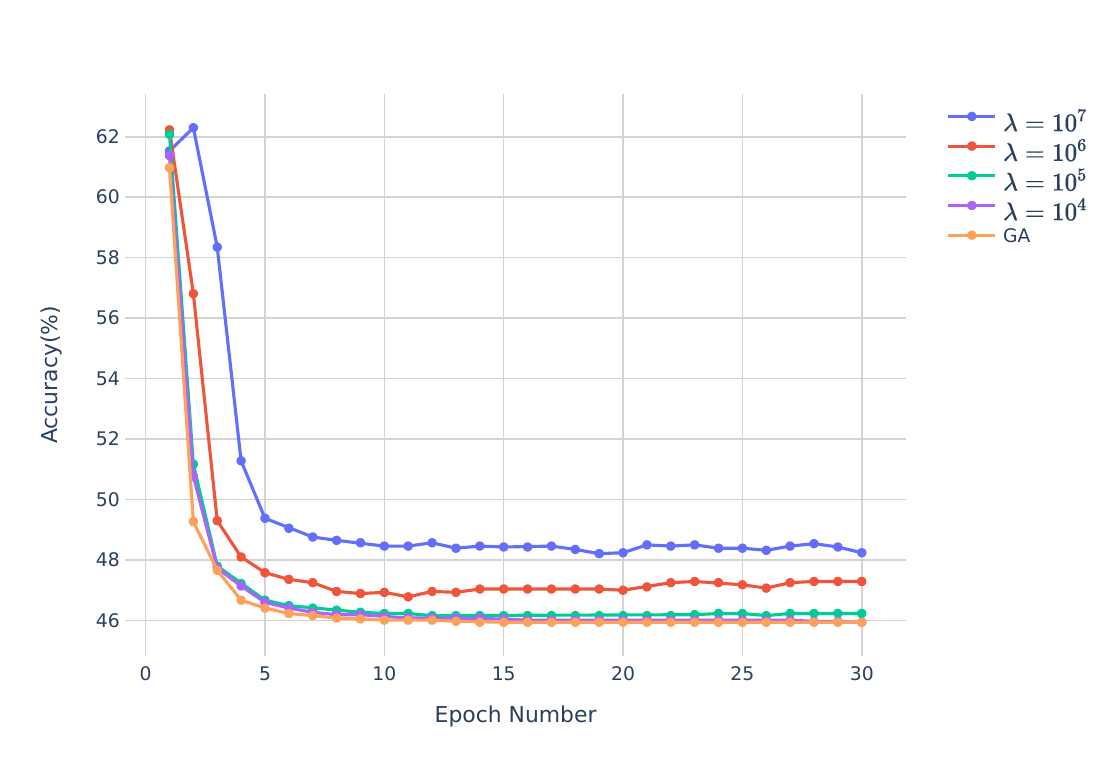}}
    %\hspace{0.1cm}
    \subfloat[ASR\label{fig:b_codebert_ga_vs_lambda_4}]{\includegraphics[scale=0.35]{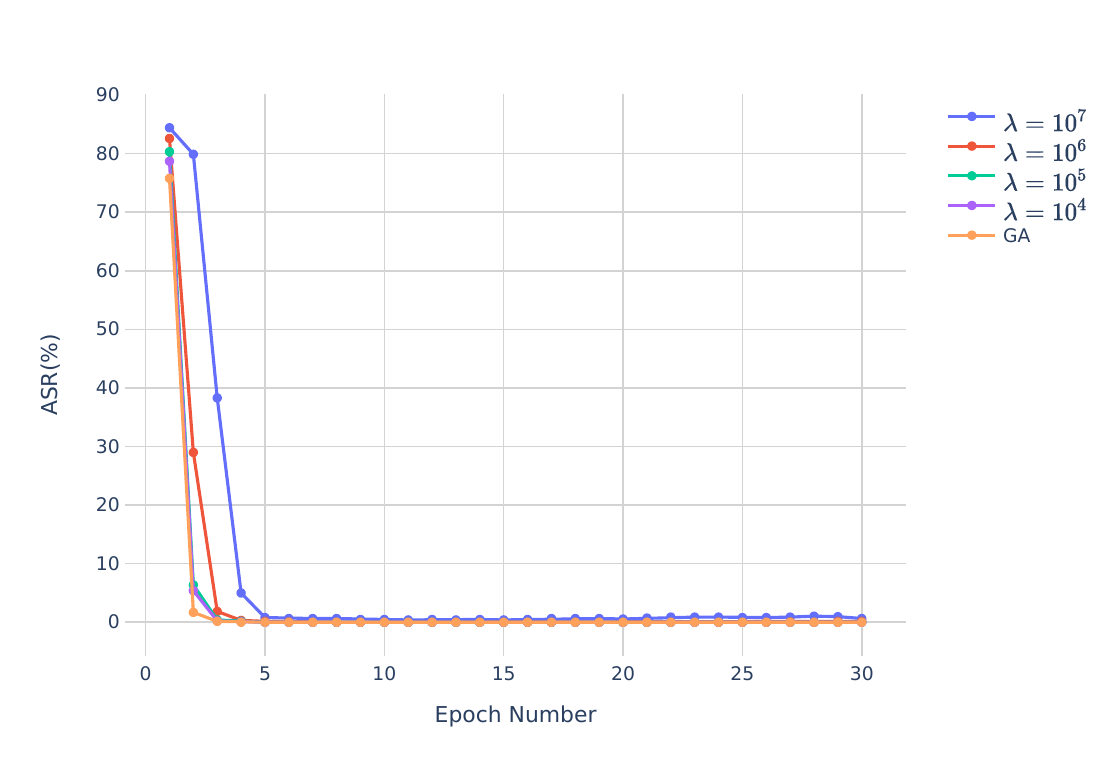}}
    \caption{\label{fig:codebert_ga_vs_lambda_4} Comparisons of Accuracy and ASR for batch size 4 and 30 epochs using the \tool (GA+EWC) approach and varied $\lambda$ values. The EWC term for poisonous datapoints is excluded from total loss (Model: CodeBERT, Task: Defect Detection, Dataset: Devign).}
\end{figure*}

%\FloatBarrier

\begin{figure*}[htbp]%[!htb]
    \centering
    \subfloat[Accuracy\label{fig:a_codebert_ga_vs_lambda_8}]{\includegraphics[scale=0.35]{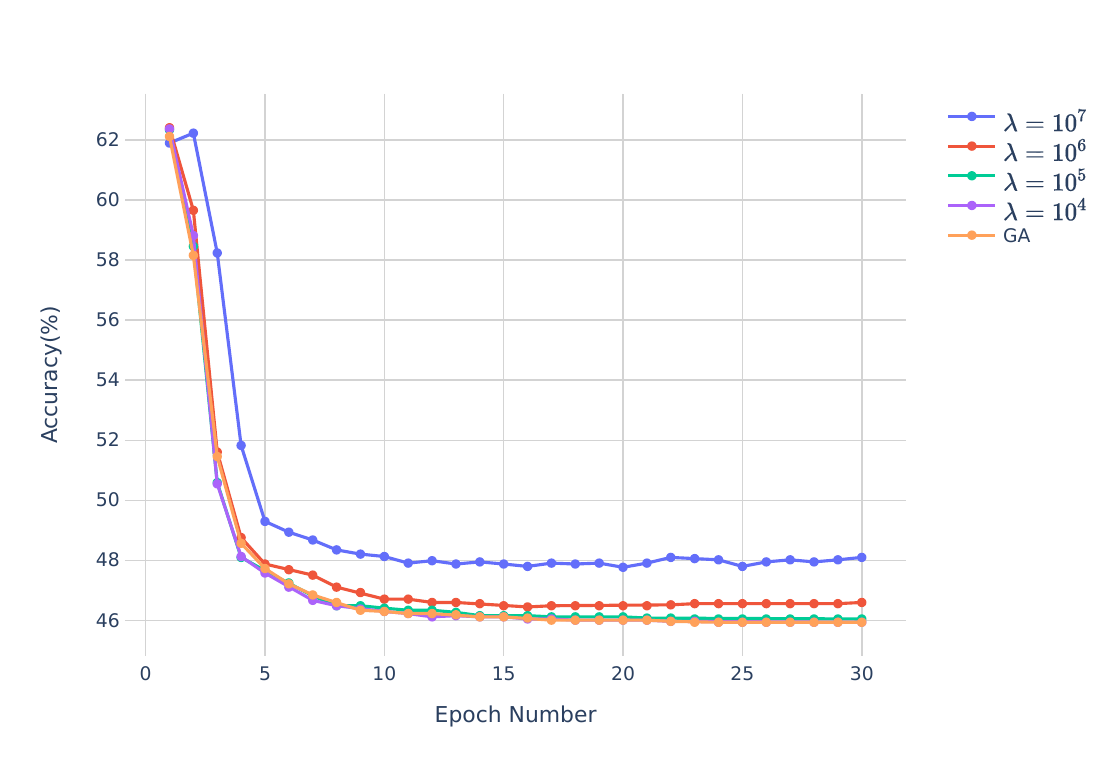}}
    %\hspace{0.1cm}
    \subfloat[ASR\label{fig:b_codebert_ga_vs_lambda_8}]{\includegraphics[scale=0.35]{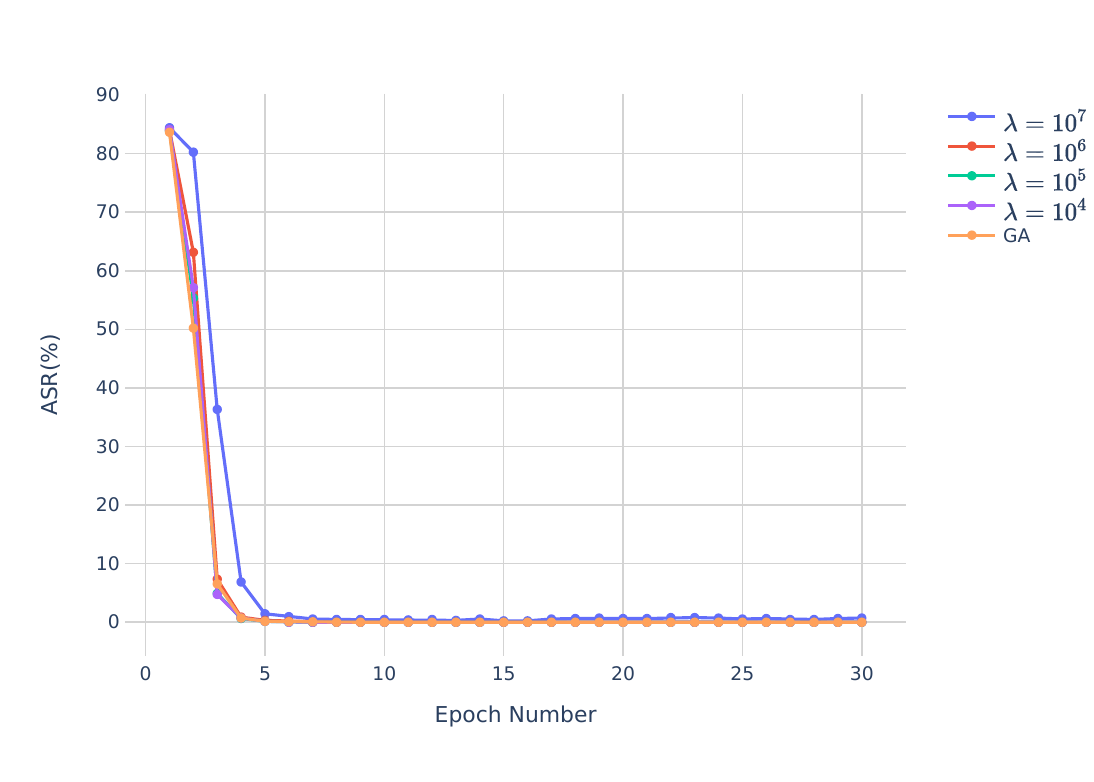}}
    \caption{\label{fig:codebert_ga_vs_lambda_8} Comparisons of Accuracy and ASR for batch size 8 and 30 epochs using the \tool (GA+EWC) approach and varied $\lambda$ values. The EWC term for poisonous datapoints is excluded from total loss (Model: CodeBERT, Task: Defect Detection, Dataset: Devign).}
\end{figure*}

%\FloatBarrier

\begin{figure*}[htbp]%[!htbp]
    \centering
    \subfloat[Accuracy\label{fig:a_codebert_ga_vs_lambda_16}]{\includegraphics[scale=0.35]{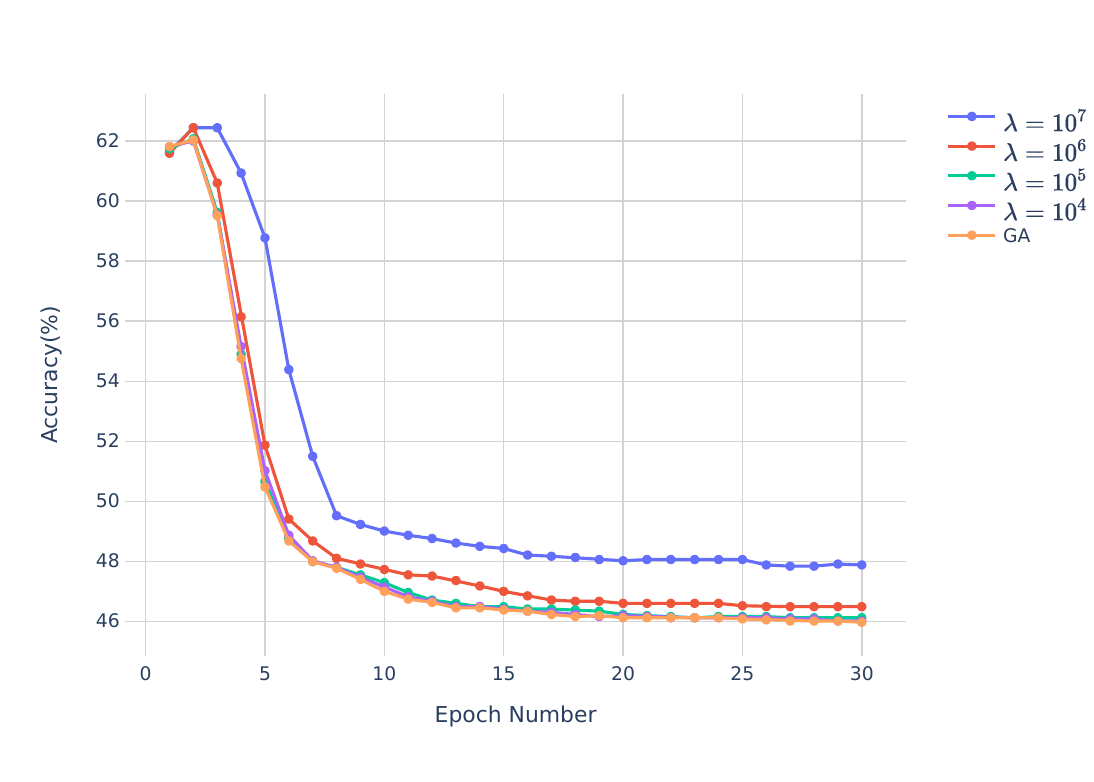}}
    %\hspace{0.1cm}
    \subfloat[ASR\label{fig:b_codebert_ga_vs_lambda_16}]{\includegraphics[scale=0.35]{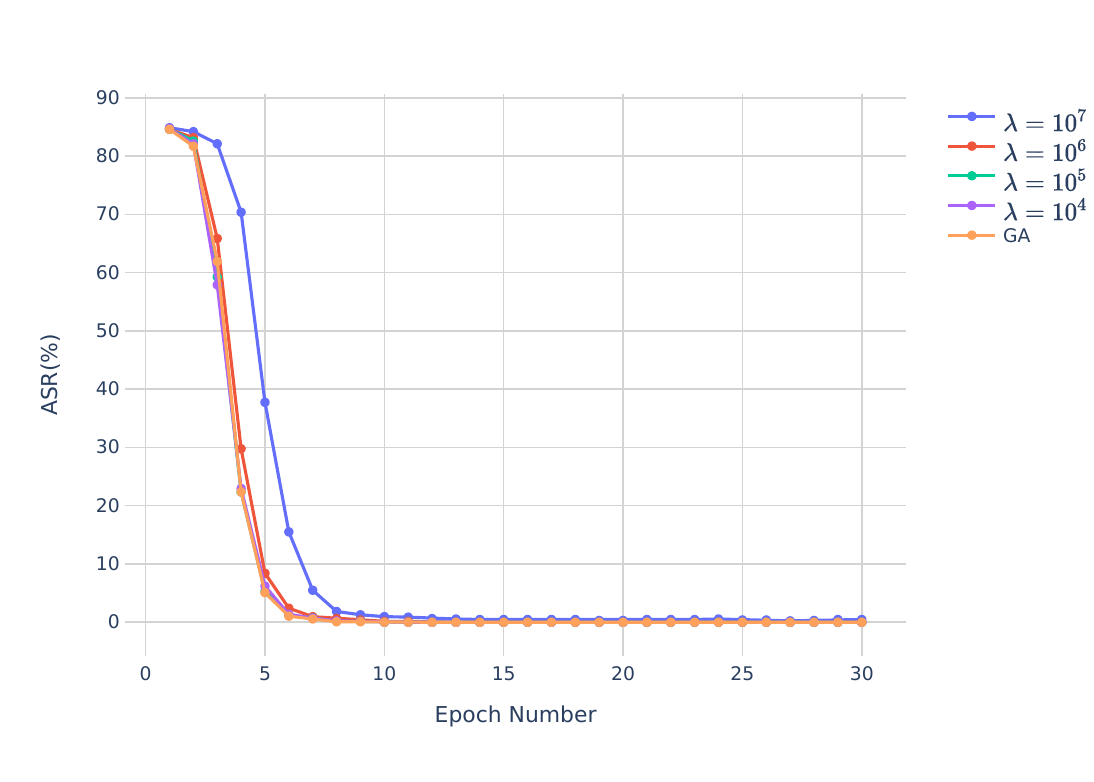}}
    \caption{\label{fig:codebert_ga_vs_lambda_16} Comparisons of Accuracy and ASR for batch size 16 and 30 epochs using the \tool (GA+EWC) approach and varied $\lambda$ values. The EWC term for poisonous datapoints is excluded from total loss (Model: CodeBERT, Task: Defect Detection, Dataset: Devign).}
\end{figure*}

%\FloatBarrier

\begin{figure*}[htbp]%[!htbp]
    \centering
    \subfloat[Accuracy\label{fig:a_codebert_ga_vs_lambda_32}]{\includegraphics[scale=0.35]{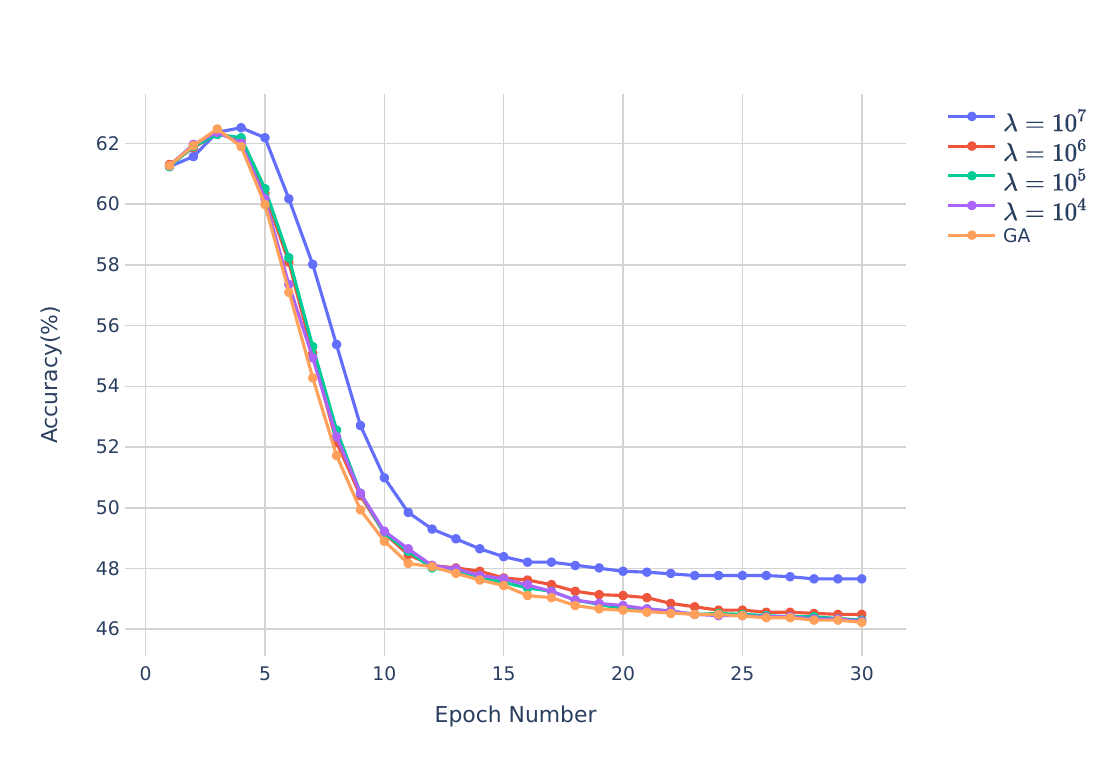}}
    %\hspace{0.1cm}
    \subfloat[ASR\label{fig:b_codebert_ga_vs_lambda_32}]{\includegraphics[scale=0.35]{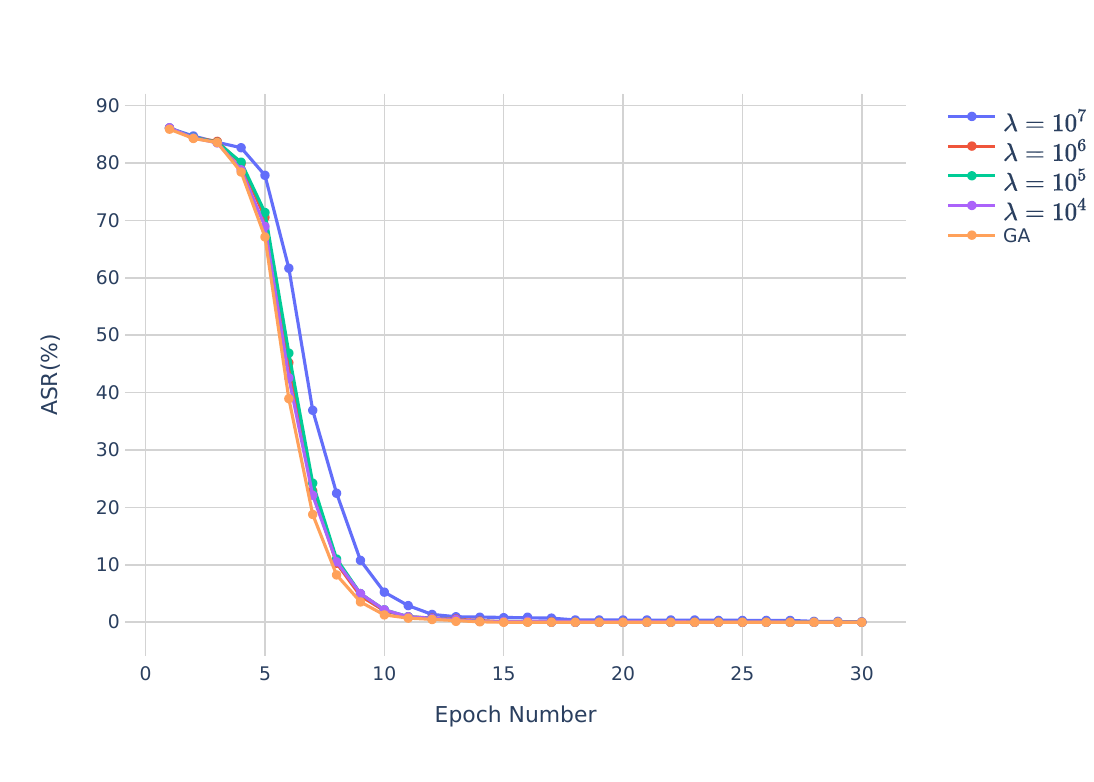}}
    \caption{\label{fig:codebert_ga_vs_lambda_32} Comparisons of Accuracy and ASR for batch size 32 and 30 epochs using the \tool (GA+EWC) approach and varied $\lambda$ values. The EWC Term for poisonous datapoints is excluded from total loss  (Model: CodeBERT, Task: Defect Detection, Dataset: Devign).}
\end{figure*}

%\FloatBarrier

\begin{figure*}[htbp]%[!htbp]
    \centering
    \subfloat[Accuracy\label{fig:a_codebert_ga_vs_lambda_64}]{\includegraphics[scale=0.35]{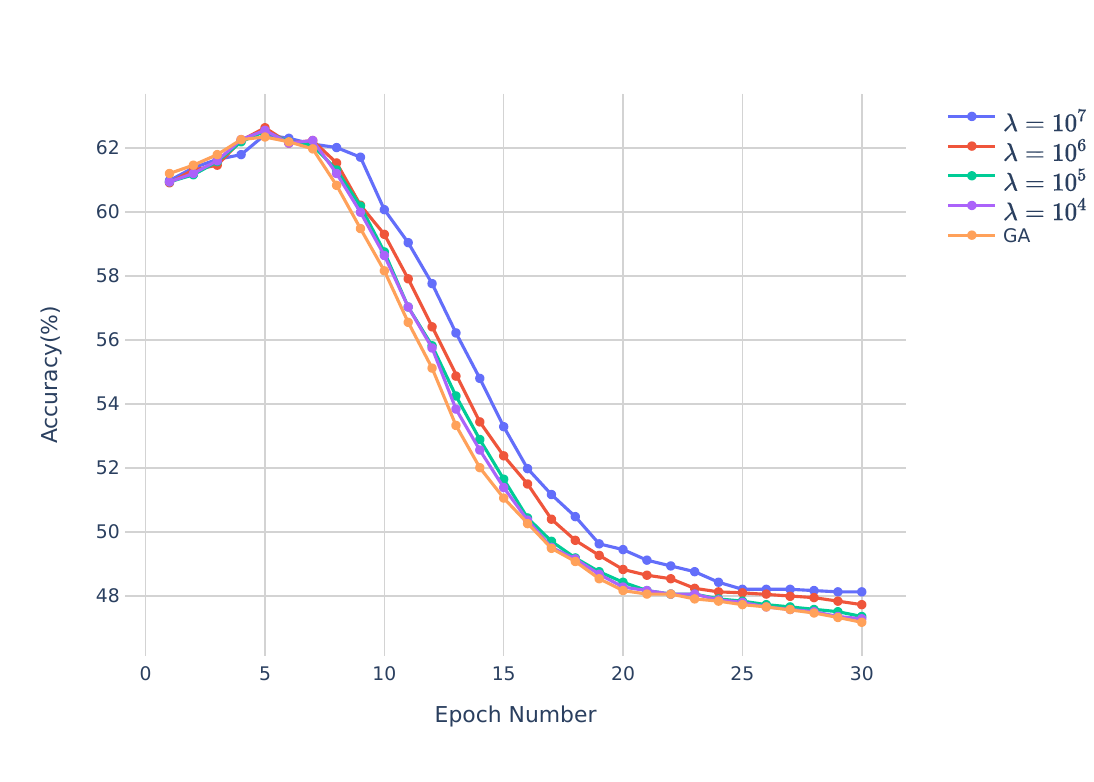}}
    %\hspace{0.1cm}
    \subfloat[ASR\label{fig:b_codebert_ga_vs_lambda_64}]{\includegraphics[scale=0.35]{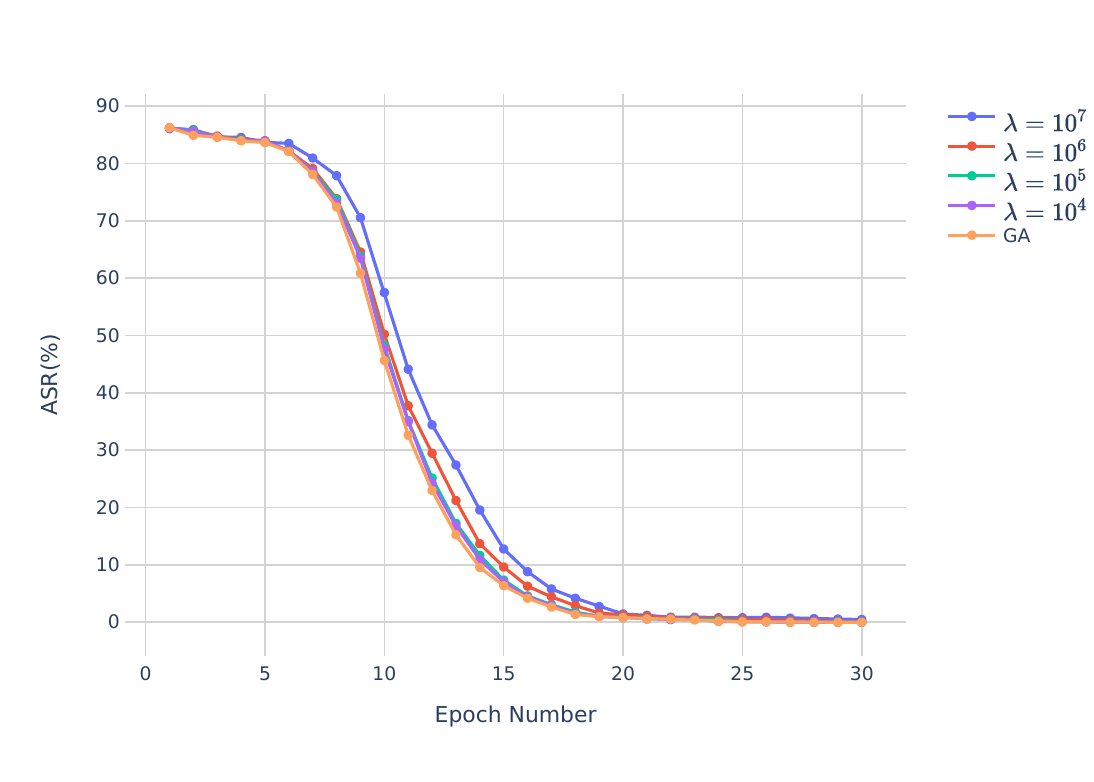}}
    \caption{\label{fig:codebert_ga_vs_lambda_64} Comparisons of Accuracy and ASR for batch size 64 and 30 epochs using the \tool (GA+EWC) approach and varied $\lambda$ values. The EWC term for poisonous datapoints is excluded from total loss (Model: CodeBERT, Task: Defect Detection, Dataset: Devign).}
\end{figure*}

%% file: disc-concl.tex
\section{Discussion and Concluding Remarks}
\label{sec:disc-concl}

From our exploration, we found unlearning in large language models, particularly when combined with EWC regularization as deployed in our new approach \tool, is effective in mitigating the impacts of trojans. The process is not only capable of reducing the ASR but also of maintaining or even improving the model's accuracy, especially when using higher $\lambda$ values and larger batch sizes. This demonstrates the potential of EWC-enhanced unlearning methods to robustly defend against trojans while preserving model performance. With regard to the comparison between unlearning in Text-LLMs and Code-LLMs, in our exploration, we found the unlearning of trojans performance is better for BERT than for CodeBERT -- in particular, the performance of CodeBERT degrades significantly in the unlearning process. This may be attributed to the inherent difference in the datasets used to obtain the pretrained versions of the two models. Code is more formal and has stricter rules to observe. Furthermore, code datasets are not as well established as natural language datasets, and thus are not as much vast and diverse, which may explain why CodeBERT finds it difficult to recover its accuracy after unlearning some poisonous data points. In the future, we look forward to making further investigations into the reasons behind the difference between unlearning performance of Text-LLMs and Code-LLMs.

%% file: main.bbl

\begin{thebibliography}{20}

%%% the Web address.

\ifx \showCODEN    \undefined \def \showCODEN     #1{\unskip}     \fi
\ifx \showDOI      \undefined \def \showDOI       #1{#1}\fi
\ifx \showISBNx    \undefined \def \showISBNx     #1{\unskip}     \fi
\ifx \showISBNxiii \undefined \def \showISBNxiii  #1{\unskip}     \fi
\ifx \showISSN     \undefined \def \showISSN      #1{\unskip}     \fi
\ifx \showLCCN     \undefined \def \showLCCN      #1{\unskip}     \fi
\ifx \shownote     \undefined \def \shownote      #1{#1}          \fi
\ifx \showarticletitle \undefined \def \showarticletitle #1{#1}   \fi
\ifx \showURL      \undefined \def \showURL       {\relax}        \fi
% invisible to TeX
\providecommand\bibfield[2]{#2}
\providecommand\bibinfo[2]{#2}
\providecommand\natexlab[1]{#1}
\providecommand\showeprint[2][]{arXiv:#2}

\bibitem[Bourtoule et~al\mbox{.}(2021)]%
        {bourtoule2021machine}
\bibfield{author}{\bibinfo{person}{Lucas Bourtoule}, \bibinfo{person}{Varun Chandrasekaran}, \bibinfo{person}{Christopher~A Choquette-Choo}, \bibinfo{person}{Hengrui Jia}, \bibinfo{person}{Adelin Travers}, \bibinfo{person}{Baiwu Zhang}, \bibinfo{person}{David Lie}, {and} \bibinfo{person}{Nicolas Papernot}.} \bibinfo{year}{2021}\natexlab{}.
\newblock \showarticletitle{Machine unlearning}. In \bibinfo{booktitle}{\emph{2021 IEEE Symposium on Security and Privacy (SP)}}. IEEE, \bibinfo{pages}{141--159}.
\newblock


\bibitem[Cao and Yang(2015)]%
        {cao2015towards}
\bibfield{author}{\bibinfo{person}{Yinzhi Cao} {and} \bibinfo{person}{Junfeng Yang}.} \bibinfo{year}{2015}\natexlab{}.
\newblock \showarticletitle{Towards making systems forget with machine unlearning}. In \bibinfo{booktitle}{\emph{2015 IEEE symposium on security and privacy}}. IEEE, \bibinfo{pages}{463--480}.
\newblock


\bibitem[Chen et~al\mbox{.}(2021)]%
        {Chen_2021}
\bibfield{author}{\bibinfo{person}{Xiaoyi Chen}, \bibinfo{person}{Ahmed Salem}, \bibinfo{person}{Dingfan Chen}, \bibinfo{person}{Michael Backes}, \bibinfo{person}{Shiqing Ma}, \bibinfo{person}{Qingni Shen}, \bibinfo{person}{Zhonghai Wu}, {and} \bibinfo{person}{Yang Zhang}.} \bibinfo{year}{2021}\natexlab{}.
\newblock \showarticletitle{BadNL: Backdoor Attacks against NLP Models with Semantic-preserving Improvements}. In \bibinfo{booktitle}{\emph{Annual Computer Security Applications Conference}} \emph{(\bibinfo{series}{ACSAC ’21})}. \bibinfo{publisher}{ACM}.
\newblock


\bibitem[Chundawat et~al\mbox{.}(2023)]%
        {chundawat2023can}
\bibfield{author}{\bibinfo{person}{Vikram~S Chundawat}, \bibinfo{person}{Ayush~K Tarun}, \bibinfo{person}{Murari Mandal}, {and} \bibinfo{person}{Mohan Kankanhalli}.} \bibinfo{year}{2023}\natexlab{}.
\newblock \showarticletitle{Can bad teaching induce forgetting? unlearning in deep networks using an incompetent teacher}. In \bibinfo{booktitle}{\emph{Proceedings of the AAAI Conference on Artificial Intelligence}}, Vol.~\bibinfo{volume}{37}. \bibinfo{pages}{7210--7217}.
\newblock


\bibitem[Devlin et~al\mbox{.}(2019)]%
        {devlin2019bert}
\bibfield{author}{\bibinfo{person}{Jacob Devlin}, \bibinfo{person}{Ming-Wei Chang}, \bibinfo{person}{Kenton Lee}, {and} \bibinfo{person}{Kristina Toutanova}.} \bibinfo{year}{2019}\natexlab{}.
\newblock \bibinfo{title}{BERT: Pre-training of Deep Bidirectional Transformers for Language Understanding}.
\newblock
\newblock
\showeprint[arxiv]{1810.04805}~[cs.CL]


\bibitem[Feng et~al\mbox{.}(2020)]%
        {feng2020codebert}
\bibfield{author}{\bibinfo{person}{Zhangyin Feng}, \bibinfo{person}{Daya Guo}, \bibinfo{person}{Duyu Tang}, \bibinfo{person}{Nan Duan}, \bibinfo{person}{Xiaocheng Feng}, \bibinfo{person}{Ming Gong}, \bibinfo{person}{Linjun Shou}, \bibinfo{person}{Bing Qin}, \bibinfo{person}{Ting Liu}, \bibinfo{person}{Daxin Jiang}, {and} \bibinfo{person}{Ming Zhou}.} \bibinfo{year}{2020}\natexlab{}.
\newblock \bibinfo{title}{CodeBERT: A Pre-Trained Model for Programming and Natural Languages}.
\newblock
\newblock
\showeprint[arxiv]{2002.08155}~[cs.CL]


\bibitem[Gao et~al\mbox{.}(2023)]%
        {gao2023keeping}
\bibfield{author}{\bibinfo{person}{Shuzheng Gao}, \bibinfo{person}{Hongyu Zhang}, \bibinfo{person}{Cuiyun Gao}, {and} \bibinfo{person}{Chaozheng Wang}.} \bibinfo{year}{2023}\natexlab{}.
\newblock \bibinfo{title}{Keeping Pace with Ever-Increasing Data: Towards Continual Learning of Code Intelligence Models}.
\newblock
\newblock
\showeprint[arxiv]{2302.03482}~[cs.SE]


\bibitem[Ginart et~al\mbox{.}(2019)]%
        {ginart2019making}
\bibfield{author}{\bibinfo{person}{Antonio Ginart}, \bibinfo{person}{Melody Guan}, \bibinfo{person}{Gregory Valiant}, {and} \bibinfo{person}{James~Y Zou}.} \bibinfo{year}{2019}\natexlab{}.
\newblock \showarticletitle{Making ai forget you: Data deletion in machine learning}.
\newblock \bibinfo{journal}{\emph{Advances in neural information processing systems}}  \bibinfo{volume}{32} (\bibinfo{year}{2019}).
\newblock


\bibitem[Golatkar et~al\mbox{.}(2020)]%
        {golatkar2020eternal}
\bibfield{author}{\bibinfo{person}{Aditya Golatkar}, \bibinfo{person}{Alessandro Achille}, {and} \bibinfo{person}{Stefano Soatto}.} \bibinfo{year}{2020}\natexlab{}.
\newblock \showarticletitle{Eternal sunshine of the spotless net: Selective forgetting in deep networks}. In \bibinfo{booktitle}{\emph{Proceedings of the IEEE/CVF Conference on Computer Vision and Pattern Recognition}}. \bibinfo{pages}{9304--9312}.
\newblock


\bibitem[Guo et~al\mbox{.}(2019)]%
        {guo2019certified}
\bibfield{author}{\bibinfo{person}{Chuan Guo}, \bibinfo{person}{Tom Goldstein}, \bibinfo{person}{Awni Hannun}, {and} \bibinfo{person}{Laurens Van Der~Maaten}.} \bibinfo{year}{2019}\natexlab{}.
\newblock \showarticletitle{Certified data removal from machine learning models}.
\newblock \bibinfo{journal}{\emph{arXiv preprint arXiv:1911.03030}} (\bibinfo{year}{2019}).
\newblock


\bibitem[Hussain et~al\mbox{.}(2024)]%
        {hussain2024trojanslargelanguagemodels}
\bibfield{author}{\bibinfo{person}{Aftab Hussain}, \bibinfo{person}{Md~Rafiqul~Islam Rabin}, \bibinfo{person}{Toufique Ahmed}, \bibinfo{person}{Bowen Xu}, \bibinfo{person}{Premkumar Devanbu}, {and} \bibinfo{person}{Mohammad~Amin Alipour}.} \bibinfo{year}{2024}\natexlab{}.
\newblock \bibinfo{title}{Trojans in Large Language Models of Code: A Critical Review through a Trigger-Based Taxonomy}.
\newblock
\newblock
\showeprint[arxiv]{2405.02828}~[cs.SE]
\urldef\tempurl%
\url{https://arxiv.org/abs/2405.02828}
\showURL{%
\tempurl}


\bibitem[Hussain et~al\mbox{.}(2023)]%
        {hussain2023trojanedcm}
\bibfield{author}{\bibinfo{person}{Aftab Hussain}, \bibinfo{person}{Md~Rafiqul~Islam Rabin}, {and} \bibinfo{person}{Mohammad~Amin Alipour}.} \bibinfo{year}{2023}\natexlab{}.
\newblock \bibinfo{title}{TrojanedCM: A Repository of Trojaned Large Language Models of Code}.
\newblock
\newblock
\showeprint[arxiv]{2311.14850}~[cs.SE]


\bibitem[Jang et~al\mbox{.}(2022)]%
        {jang2022knowledge}
\bibfield{author}{\bibinfo{person}{Joel Jang}, \bibinfo{person}{Dongkeun Yoon}, \bibinfo{person}{Sohee Yang}, \bibinfo{person}{Sungmin Cha}, \bibinfo{person}{Moontae Lee}, \bibinfo{person}{Lajanugen Logeswaran}, {and} \bibinfo{person}{Minjoon Seo}.} \bibinfo{year}{2022}\natexlab{}.
\newblock \showarticletitle{Knowledge unlearning for mitigating privacy risks in language models}.
\newblock \bibinfo{journal}{\emph{arXiv preprint arXiv:2210.01504}} (\bibinfo{year}{2022}).
\newblock


\bibitem[Kirkpatrick et~al\mbox{.}(2017)]%
        {Kirkpatrick_2017}
\bibfield{author}{\bibinfo{person}{James Kirkpatrick}, \bibinfo{person}{Razvan Pascanu}, \bibinfo{person}{Neil Rabinowitz}, \bibinfo{person}{Joel Veness}, \bibinfo{person}{Guillaume Desjardins}, \bibinfo{person}{Andrei~A. Rusu}, \bibinfo{person}{Kieran Milan}, \bibinfo{person}{John Quan}, \bibinfo{person}{Tiago Ramalho}, \bibinfo{person}{Agnieszka Grabska-Barwinska}, \bibinfo{person}{Demis Hassabis}, \bibinfo{person}{Claudia Clopath}, \bibinfo{person}{Dharshan Kumaran}, {and} \bibinfo{person}{Raia Hadsell}.} \bibinfo{year}{2017}\natexlab{}.
\newblock \showarticletitle{Overcoming catastrophic forgetting in neural networks}.
\newblock \bibinfo{journal}{\emph{Proceedings of the National Academy of Sciences}} \bibinfo{volume}{114}, \bibinfo{number}{13} (\bibinfo{date}{March} \bibinfo{year}{2017}), \bibinfo{pages}{3521–3526}.
\newblock
\showISSN{1091-6490}
\urldef\tempurl%
\url{https://doi.org/10.1073/pnas.1611835114}
\showDOI{\tempurl}


\bibitem[Koh and Liang(2017)]%
        {koh2017understanding}
\bibfield{author}{\bibinfo{person}{Pang~Wei Koh} {and} \bibinfo{person}{Percy Liang}.} \bibinfo{year}{2017}\natexlab{}.
\newblock \showarticletitle{Understanding black-box predictions via influence functions}. In \bibinfo{booktitle}{\emph{International conference on machine learning}}. PMLR, \bibinfo{pages}{1885--1894}.
\newblock


\bibitem[Maas et~al\mbox{.}(2011)]%
        {maas2011learning}
\bibfield{author}{\bibinfo{person}{Andrew Maas}, \bibinfo{person}{Raymond~E Daly}, \bibinfo{person}{Peter~T Pham}, \bibinfo{person}{Dan Huang}, \bibinfo{person}{Andrew~Y Ng}, {and} \bibinfo{person}{Christopher Potts}.} \bibinfo{year}{2011}\natexlab{}.
\newblock \showarticletitle{Learning word vectors for sentiment analysis}. In \bibinfo{booktitle}{\emph{Proceedings of the 49th annual meeting of the association for computational linguistics: Human language technologies}}. \bibinfo{pages}{142--150}.
\newblock


\bibitem[Mehta et~al\mbox{.}(2022)]%
        {mehta2022deep}
\bibfield{author}{\bibinfo{person}{Ronak Mehta}, \bibinfo{person}{Sourav Pal}, \bibinfo{person}{Vikas Singh}, {and} \bibinfo{person}{Sathya~N Ravi}.} \bibinfo{year}{2022}\natexlab{}.
\newblock \showarticletitle{Deep unlearning via randomized conditionally independent hessians}. In \bibinfo{booktitle}{\emph{Proceedings of the IEEE/CVF Conference on Computer Vision and Pattern Recognition}}. \bibinfo{pages}{10422--10431}.
\newblock


\bibitem[Wang et~al\mbox{.}(2023)]%
        {wang2023kga}
\bibfield{author}{\bibinfo{person}{Lingzhi Wang}, \bibinfo{person}{Tong Chen}, \bibinfo{person}{Wei Yuan}, \bibinfo{person}{Xingshan Zeng}, \bibinfo{person}{Kam-Fai Wong}, {and} \bibinfo{person}{Hongzhi Yin}.} \bibinfo{year}{2023}\natexlab{}.
\newblock \showarticletitle{Kga: A general machine unlearning framework based on knowledge gap alignment}.
\newblock \bibinfo{journal}{\emph{arXiv preprint arXiv:2305.06535}} (\bibinfo{year}{2023}).
\newblock


\bibitem[Zhang et~al\mbox{.}(2024)]%
        {zhang2024negativepreference}
\bibfield{author}{\bibinfo{person}{Ruiqi Zhang}, \bibinfo{person}{Licong Lin}, \bibinfo{person}{Yu Bai}, {and} \bibinfo{person}{Song Mei}.} \bibinfo{year}{2024}\natexlab{}.
\newblock \bibinfo{title}{Negative Preference Optimization: From Catastrophic Collapse to Effective Unlearning}.
\newblock
\newblock
\showeprint[arxiv]{2404.05868}~[cs.LG]
\urldef\tempurl%
\url{https://arxiv.org/abs/2404.05868}
\showURL{%
\tempurl}


\bibitem[Zhou et~al\mbox{.}(2019)]%
        {zhou2019devign}
\bibfield{author}{\bibinfo{person}{Yaqin Zhou}, \bibinfo{person}{Shangqing Liu}, \bibinfo{person}{Jingkai Siow}, \bibinfo{person}{Xiaoning Du}, {and} \bibinfo{person}{Yang Liu}.} \bibinfo{year}{2019}\natexlab{}.
\newblock \bibinfo{title}{Devign: Effective Vulnerability Identification by Learning Comprehensive Program Semantics via Graph Neural Networks}.
\newblock
\newblock
\showeprint[arxiv]{1909.03496}~[cs.SE]


\end{thebibliography}
